\documentclass[global,twocolumn]{svjour} 

\usepackage{latexsym}
\usepackage{graphics}
\usepackage{subfig}
\usepackage{float}
\usepackage{fixltx2e} 
\usepackage{color}

\begin{document}
\titlerunning{\textsc{Pentatrap}}
\title{PENTATRAP: A novel cryogenic multi-Penning trap experiment for high-precision mass measurements on highly charged ions}
\author{J.~Repp\inst{1,2}\thanks{\emph{Corresponding author:} Fax: (+496221) 516-852, E-mail: julia.repp@mpi-hd.mpg.de, } \and Ch.~B\"ohm\inst{1,2,3} \and J.R.~Crespo L\'opez-Urrutia\inst{1} \and A.~D\"orr\inst{1,2} \and S.~Eliseev\inst{1} \and S.~George\inst{1}\thanks{\emph{Present address:} National Superconducting Cyclotron Laboratory, MSU, East Lansing, MI, 48824-1321, USA} \and M.~Goncharov\inst{1,2} \and Yu.N.~Novikov\inst{3,4} \and C.~Roux\inst{1,2} \and S.~Sturm\inst{1,5} \and S.~Ulmer\inst{1,2,5}\thanks{\emph{Present address:} RIKEN Advanced Science Institute, Hirosawa, Wako, Saitama 351-0198, Japan} \and K.~Blaum\inst{1,2}
}
\offprints{J.~Repp}
\institute{Max-Planck-Institut f\"ur Kernphysik, 69117 Heidelberg, Germany \and Fakult\"at f\"ur Physik und Astronomie, Ruprecht-Karls-Universit\"at, 69120 Heidelberg, Germany \and Extreme Matter Institute EMMI, Helmholtz Gemeinschaft, 64291 Darmstadt, Germany \and Petersburg Nuclear Physics Institute, 188300 Gatchina, Russia \and Johannes Gutenberg-Universit\"at Mainz, Institut f\"ur Physik, 55099 Mainz, Germany}
\maketitle

\begin{abstract}
The novel five-Penning trap mass spectrometer \textsc{Pentatrap} is developed at the Max-Planck-Insti\-tut f\"ur Kernphysik (MPIK), Heidelberg. Ions of interest are long-lived highly charged nuclides up to bare uranium. \textsc{Pentatrap} aims for an accuracy of a few parts in 10$^{12}$ for mass ratios of mass doublets.
A physics program for \textsc{Pentatrap} includes \textit{Q}-values measurements of $\beta$-transitions relevant for neutrino physics, stringent tests of quantum electrodynamics in the regime of extreme electric fields, and a test of special relativity.
Main features of \textsc{Pentatrap} are an access to a source of highly charged ions, a multi-trap configuration, simultaneous measurements of frequencies, a continuous precise monitoring of magnetic field fluctuations, a fast exchange between different ions, and a highly sensitive cryogenic non-destruc\-tive detection system.
This paper gives a motivation for the new mass spectrometer \textsc{Pentatrap}, presents its experimental setup, and describes the present status.
\end{abstract}

\section{Introduction and motivation}
\label{Introduction-and-motivation}
The mass of a nuclide is an important parameter in many fields of physics and reflects all forces acting in the nucleus \cite{Blaum2006,Franzke2008,Blaum2010}. The necessary relative uncertainty of the mass measurement depends on the physics being investigated
and ranges from $\delta m/m\approx 10^{-7}-10^{-8}$ in the field of nuclear physics \cite{Lunney2003,Blaum2003,Kankainen2010} and astrophysics \cite{Weber2008,Baruah2008,Dworschak2008,Elomaa2009}, down to the smallest possible uncertainties of better than $10^{-11}$ for tests of fundamental interactions and their symmetries \cite{Hardy2009,Kellerbauer2004,Mukherjee2004,Savard2005,Bollen2006,Eronen2006,George2007,Eronen2008,Eronen2009,Gabrielse1999,Shabaev2006,Stoehlker2008,Rainville2005}.
\\
Penning traps are nowadays the most suitable devices for high-precision mass measurements of nuclides. The mass is determined via the measurement of the free-space cyclotron frequency
\begin{equation}
\nu_c=\frac{1}{2\pi}\frac{qB}{m}
\label{eq:cyclotronfrequency}
\end{equation}
of an ion with charge-to-mass ratio $q/m$ stored in a homogeneous magnetic field $B$. Frequencies can generally be measured with very high precision. Therefore, relating a mass measurement to a frequency measurement is advantageous.
The most accurate mass measurements with a relative uncertainty below 10$^{-11}$ have been performed with light and stable atoms and molecules created inside the trap volume \cite{VanDyck2004,VanDyck2006,Rainville2004,Shi2005,Redshaw2008}.
\\
At present, direct access to highly charged stable ions and an accuracy of a few 10$^{-10}$ in the mass determination are solely realizable at the \textsc{Smiletrap} Penning trap facility \cite{Bergstroem2002,Bergstroem2003}. In the near future, the \textsc{Titan} facility aims to carry out high-precision mass measurements on highly charged radionuclides \cite{Dilling2006}. Both experiments use the time-of-flight ion cyclotron resonance measurement technique \cite{Graeff1980} and presently cannot provide access to heavy highly charged ions.
The \textsc{Pentatrap} experiment is currently being built at the Max-Planck-Insti\-tut f\"ur Kernphysik in Heidelberg, in order to access stable and radioactive highly charged heavy nuclides up to uranium, and to measure their mass ratios with an accuracy of few parts in 10$^{12}$. To enable these ultra-high accuracies, \textsc{Pentatrap} employs a non-destructive image-current detection technique \cite{Wineland1975}, using a highly sensitive cryogenic detection system and a stack of five Penning traps.
\\
The mass ratios measured at \textsc{Pentatrap} will contribute to a determination of neutrino properties \cite{Blaum2010,Eliseev2011}. To this end, the mass difference between the initial and final nucleus of radioactive processes of interest must be determined. Examples are the electron capture process\linebreak $^{163}$Ho$\,$+$\,$ e$^-$$\,\rightarrow\,$$^{163}$Dy$\,$+$\,\nu_\mathrm{e}$ \cite{Rujula1982} and the $\beta$-decay\linebreak $^{187}$Re$\,\rightarrow\,$$^{187}$Os$\,$+$\,$e$^-$+$\,\bar{\nu}_\mathrm{e}$ \cite{Ferri2009}. 
In combination with cryogenic microcalorimetry, these measurements might enable one to probe the neutrino mass at a sub-eV level \cite{Otten2008}. 
Another application for \textsc{Pentatrap} are stringent tests of quantum electrodynamics in the regime of extreme fields. Here, the determination of the binding energy of the last remaining $1s$ electron in highly charged $^{208}$Pb is planned at the level of better than 1 eV, which would exceed the present precision of X-ray spectroscopy \cite{Stoehlker2006} by at least a factor of three.
The high-precision mass measurements of \textsc{Pentatrap} can also contribute to tests of fundamental symmetries and constants, e.g., to a direct test of the mass-energy relationship of the relativistic theory. Here, highly accurate mass measurements of atomic mass differences of nuclides involved in neutron capture processes combined with $\gamma$-ray spectroscopy allow a determination of nuclear binding energies \cite{Rainville2005}.\\ 
This article is structured as follows: The basic principles of Penning trap mass spectrometry will be summarized in Sec.~\ref{Theory}. In Sec.~\ref{Expsetup} the novel design of the \textsc{Pentatrap} experiment and its specific features, followed by the ion detection system in Sec.~\ref{Iondetection}, will be presented. Finally, Sec.~\ref{measurementprocedure} deals with the exploration of different measurement schemes.

\section{Principles of high-precision Penning trap mass spectrometry}
\label{Theory}
\begin{figure*}
\resizebox{0.9\textwidth}{!}{
  \includegraphics{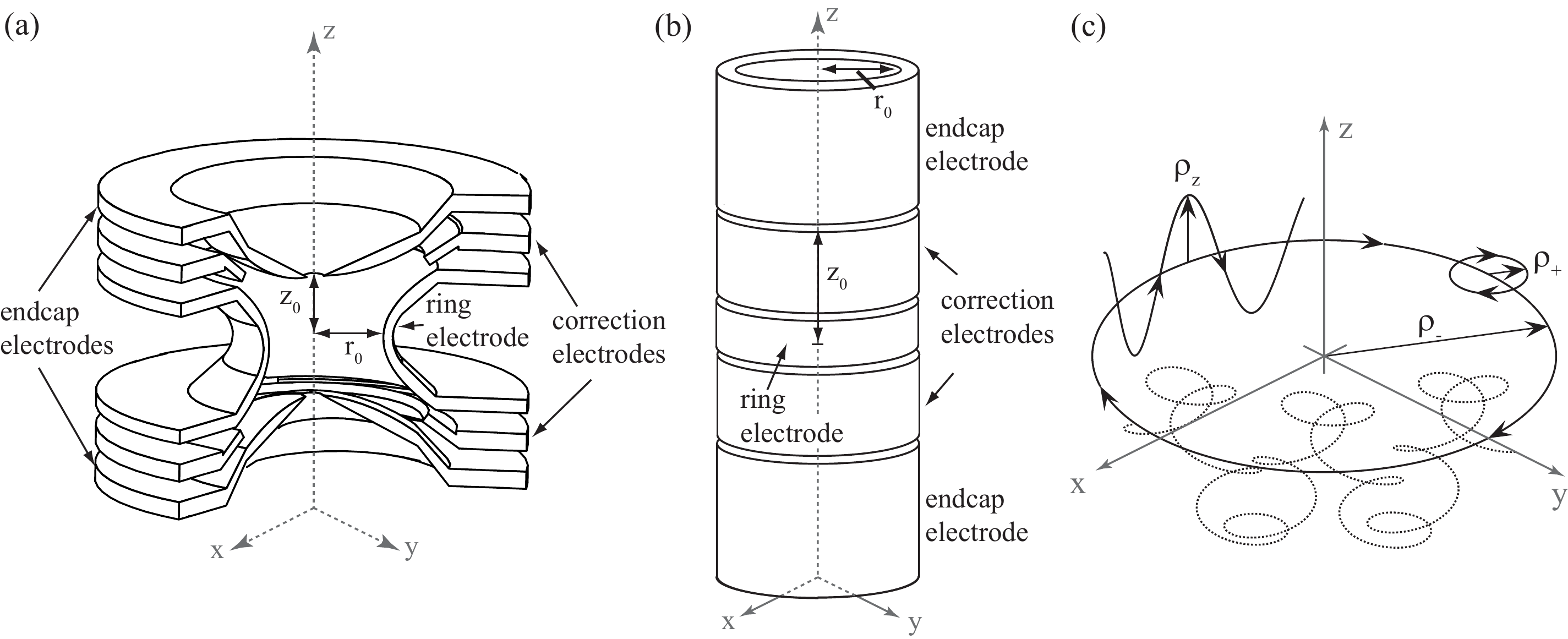}}
\caption{Sketch of a five-electrode hyperboloidal (a) and cylindrical (b) Penning trap consisting of ring, correction and endcap electrodes. (c) shows the complete ion motion in a Penning trap (dashed line), which is a superposition of three modes with amplitudes $\rho_+$, $\rho_-$ and $\rho_z$.}
\label{fig:CylindricalTraps} 
\end{figure*}

The storage of a charged particle in an ideal Penning trap is realized through a superposition of a strong homogeneous magnetic field and a weak electrostatic field for radial and axial confinement, respectively \cite{Brown1986}.
The three-dimensional quadrupolar electrostatic potential near the trap center can be created by electrodes formed either as hyperbolas of revolution (Fig.~\ref{fig:CylindricalTraps}(a)) or 
as a cylindrical trap structure (Fig.~\ref{fig:CylindricalTraps}(b)). Cylindrical structures are often used due to their easier manufacturing and free access to the trap center for particle loading \cite{Gabrielse1989}. The resulting ion motion is a superposition of three modes (Fig.~\ref{fig:CylindricalTraps}(c)). In the axial direction the ion performs a harmonic motion with the frequency 
\begin{equation}
\nu_z=\frac{1}{2\pi}\sqrt{\frac{q}{m}\frac{U_0}{d_0^2}},
\label{eq:axialfrequency}
\end{equation}
with $U_0$ the voltage applied to the ring electrode (see Fig.~\ref{fig:CylindricalTraps}), and $d_0$ a geometrical parameter characterizing the trap size \cite{Gabrielse1989}.
The radial motional frequencies called modified cyclotron frequency $\nu_+$ and magnetron frequency $\nu_-$ are given by
\begin{equation}
\nu_\pm=\frac{\nu_c}{2}\pm\sqrt{\frac{\nu_c^2}{4}-\frac{\nu_z^2}{2}}.
\label{eq:radialfrequencies}
\end{equation}
The invariance theorem \cite{Gabrielse2009} gives the relation between the free-space cyclotron frequency and the eigenfrequencies, which is
\begin{equation}
\nu_c^2=\nu_z^2+\nu_+^2+\nu_-^2,
\label{eq:invariancetheorem}
\end{equation}
and is even valid in the case of harmonic imperfections of the electrical field and misalignments between the electric and magnetic field axis \cite{Brown1986}.
A measurement of all three ion frequencies thus results in a determination of the ion's mass (see Eq.~(\ref{eq:cyclotronfrequency})). In the case of determining mass ratios, the strength of the magnetic field ideally cancels out by alternatingly measuring the cyclotron frequencies. 
\\
In a real trap, the ion's eigenfrequencies can be affected by, e.g., anharmonicity of the electric potential or inhomogeneity of the magnetic field \cite{Brown1986}, which leads to an increase of the uncertainty of mass measurements. 
Using a five-electrode trap structure with suitable geometric parameters (see Fig.~\ref{fig:CylindricalTraps}), a sufficient suppression of anharmonic terms in the electric potential around the trap center can be achieved. Unlike the electrostatic potential, higher order terms in the Legendre polynomial expansion of the magnetic field, especially the second order term responsible for a magnetic bottle, often cannot be tuned as easily. Therefore, only superconducting magnets with a high spatial homogeneity are used in high-precision Penning-trap mass spectrometry.
Moreover, the real magnetic field shows a temporal drift and short-term fluctuations \cite{Liu2010,Droese2011}. Thus, a magnetic field stabilization \cite{VanDyck1999} and a fast measurement of the frequency ratio is essential to gain very high precision.

\section{Experimental setup}
\label{Expsetup}
\begin{figure}
	\resizebox{.49\textwidth}{!}{
		\includegraphics{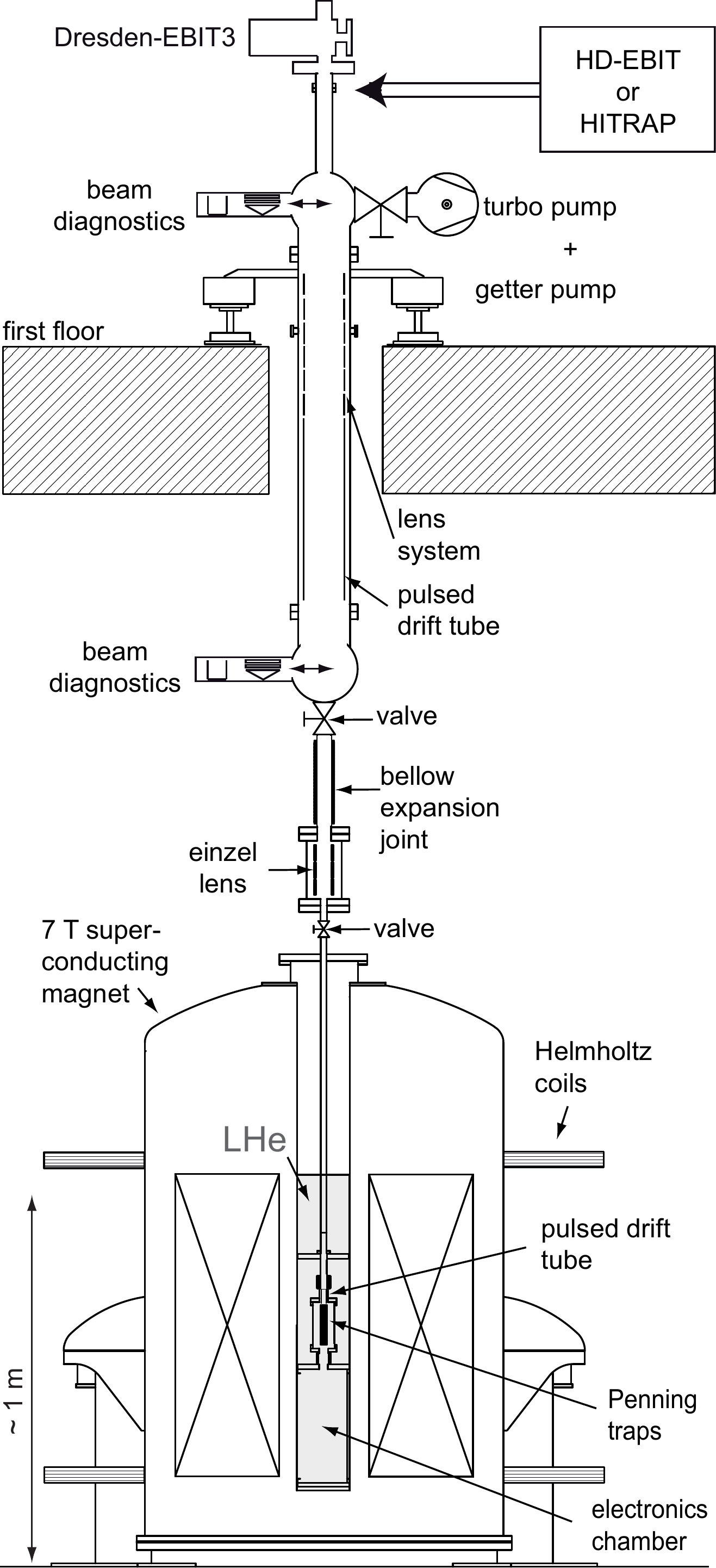}}
	\caption{Layout of the experimental setup. Ions will be created by either a commercial electron beam ion trap (EBIT) or the Heidelberg-EBIT. At GSI, ions might be obtained from the \textsc{Hitrap} facility \cite{Kluge2008}. The ions will pass the beamline towards the superconducting magnet, which is situated in a temperature-stabilized and pressure-regulated room in the basement. The beam diagnostic elements and the electrostatic lens system are shown. The cryogenic chambers for the trap tower and the detection electronics are placed inside the cold bore of the magnet.}
	\label{fig:PENTATRAPConcept}
\end{figure}

\begin{figure}
	\resizebox{.46\textwidth}{!}{
		\includegraphics{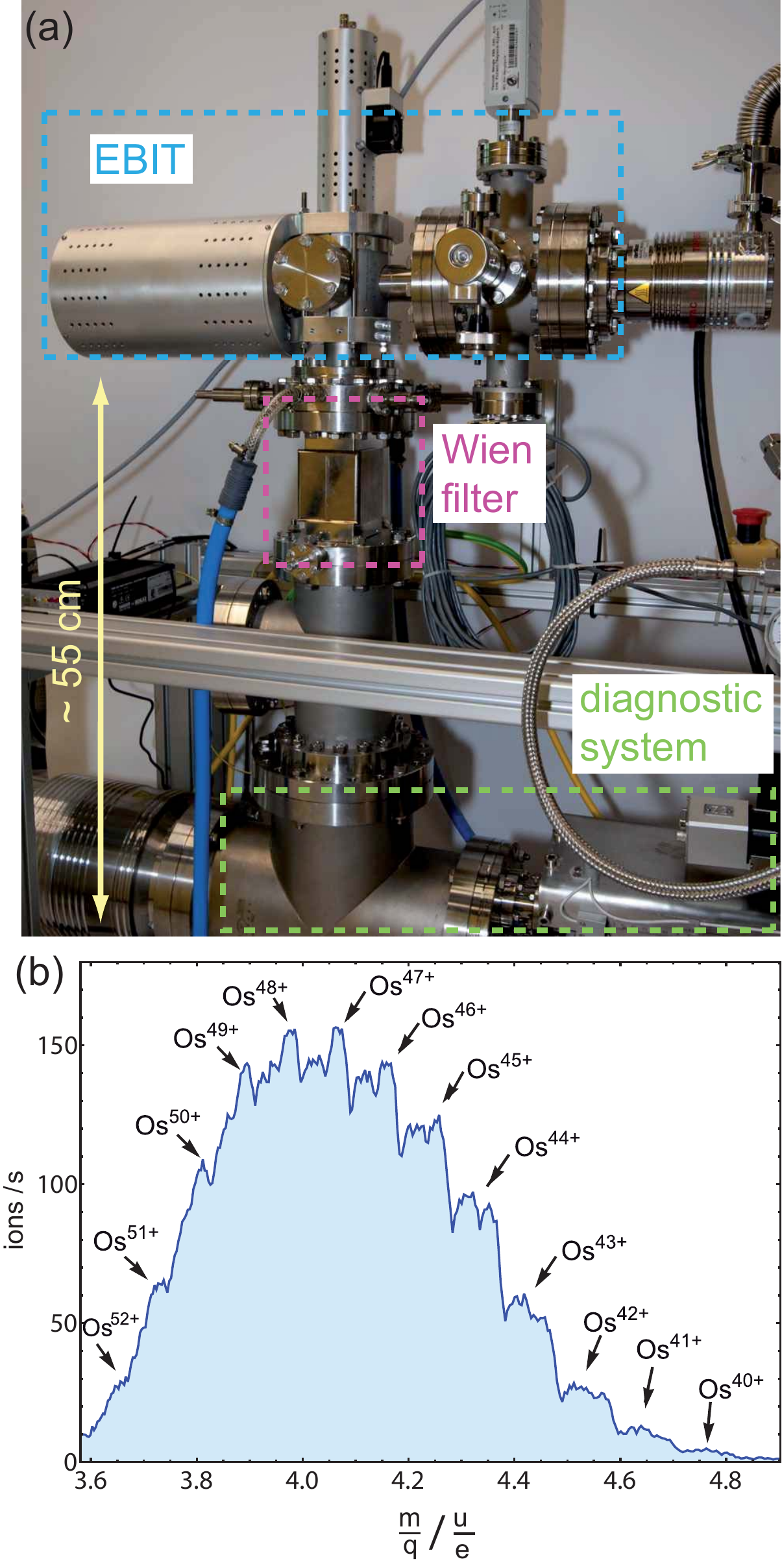}}
	\caption{(Color online)(a) Photograph of the Dresden-EBIT3 setup. (b) Count rate of Os-ions extracted from the Dresden-EBIT3 and detected with a multi-channel plate detector for different mass-to-charge ratios. The setup was operated with a trapping time of 800$\,$ms, an electron current and energy of 30$\,$mA and 7.9$\,$keV, respectively, a trap depth of \mbox{$-100\,$V} and an ion energy of 4.9$\,$kV$\cdot$q. Osmium ions as well as various types of molecular fragments are produced by sublimation of osmocen Os(C$_5$H$_5$)$_2$ before the gas is fed into the trapping volume and ionized. The osmium in the compound has the most frequent natural abundances of 40.78\% ($^{192}$Os), 26.26\% ($^{190}$Os), 16.15\% ($^{189}$Os), 13.24\% ($^{188}$Os) and 1.96\% ($^{187}$Os). In the spectrum, the resolving power was not high enough to resolve the different Os isotopes. It is possible to create charge states up to 52+. Later enriched osmocen will be used to produce more $^{187}$Os ions.}
	\label{fig:EBIT3}
\end{figure}

\begin{figure}
	\resizebox{.48\textwidth}{!}{
		\includegraphics{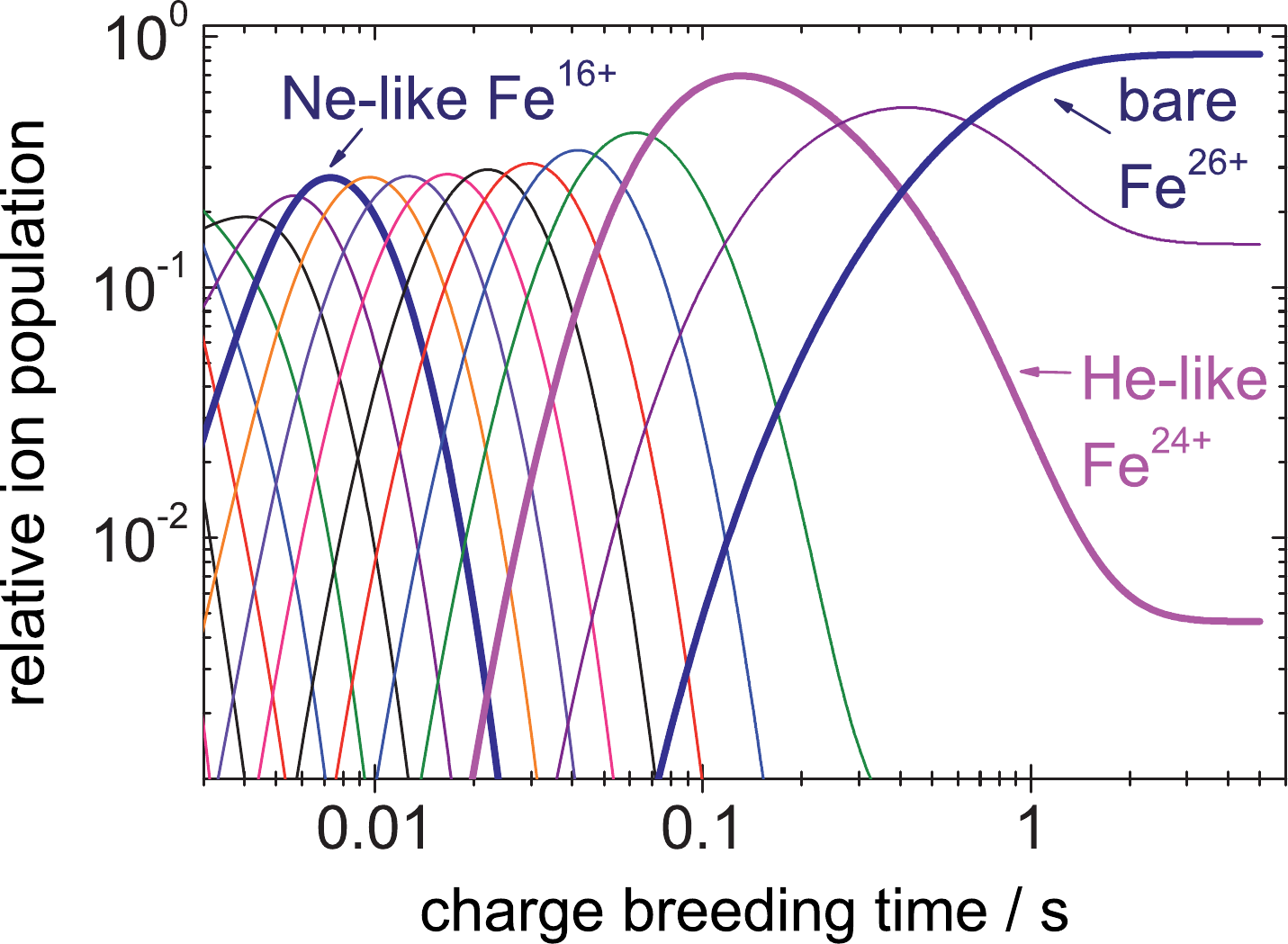}}
	\caption{(Color online) Calculated fraction of Fe ions in charge state $n+$ versus charge breeding time after injection into the Heidelberg-EBIT. An electron beam current of 500$\,$mA and an electron/ion overlap of 0.1 are assumed. The electron beam energy is set to 20$\,$keV. For charge breeding times greater than 2$\,$s, 85$\,\%$ of the Fe ions are bare.}
	\label{fig:EBIT}
\end{figure}

\begin{figure}
	\resizebox{.49\textwidth}{!}{
		\includegraphics{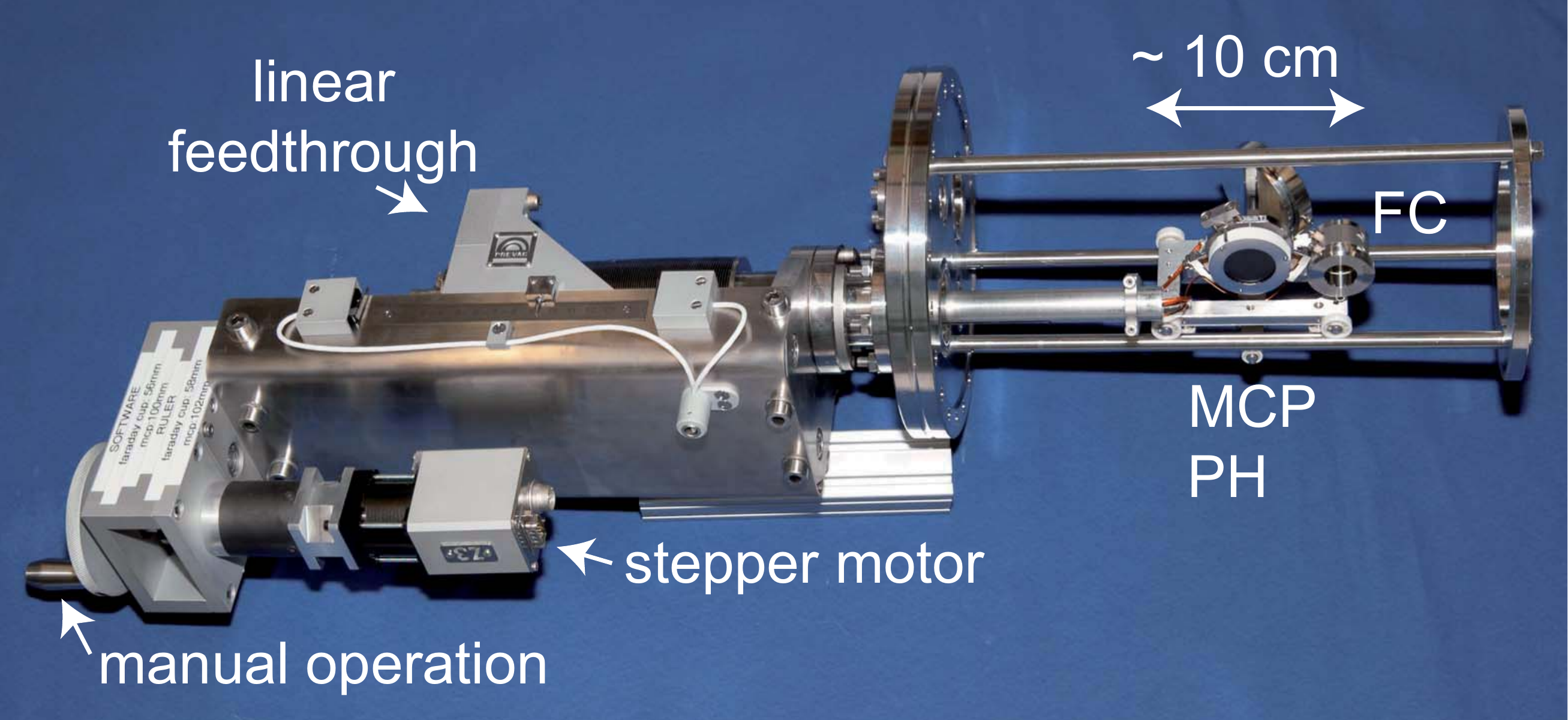}}
		\caption{(Color online) Beam diagnostic system consisting of a Faraday cup (FC) and a multi-channel plate (MCP) detector with attached phosphor screen (PH).}
			\label{fig:DiagnoseSystem}
\end{figure}

\begin{figure*}
	\resizebox{.8\textwidth}{!}{
		\includegraphics{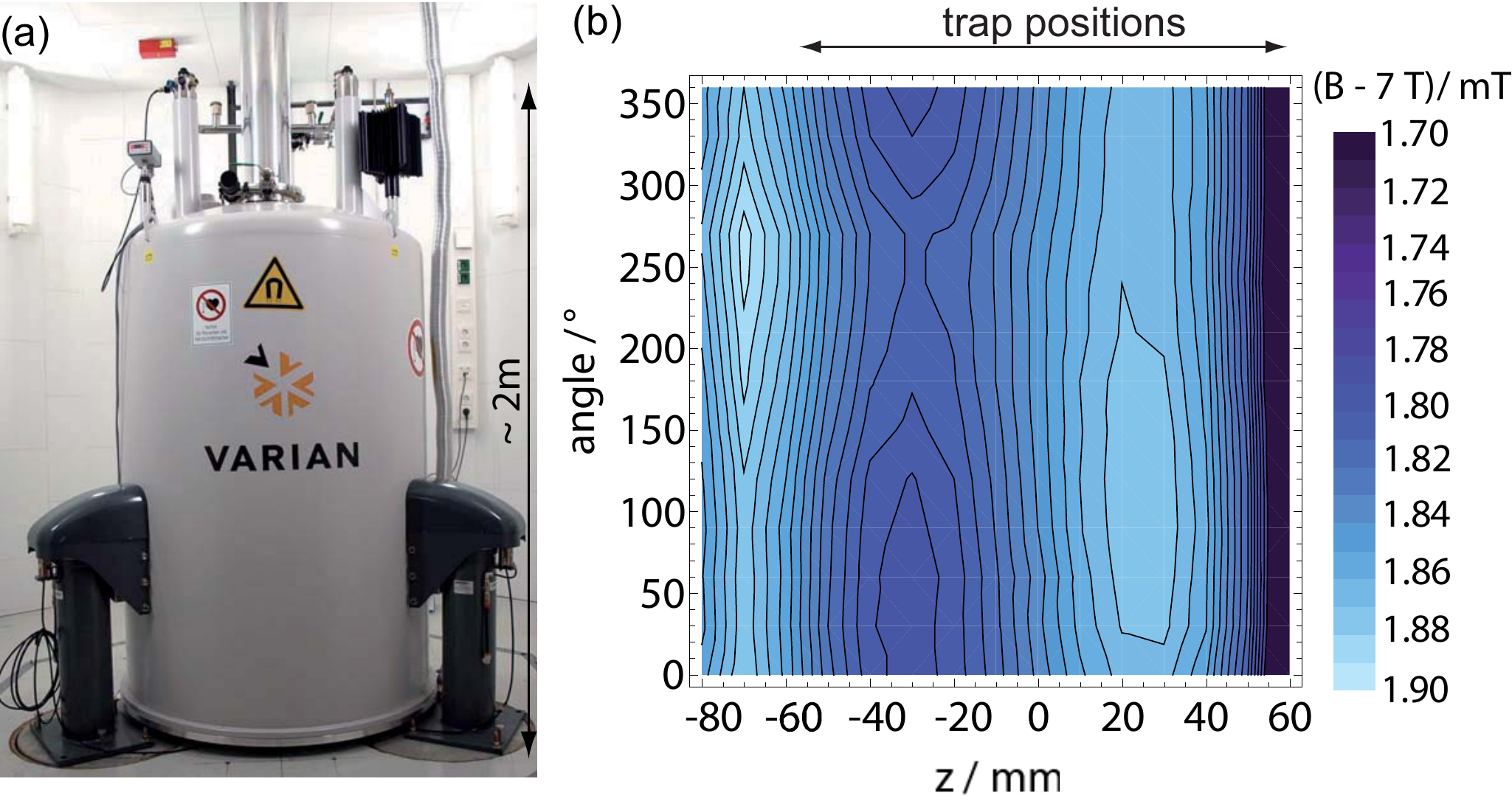}}
	\caption{(Color online)(a) Photograph of the \textsc{Pentatrap} magnet and its installation in the temperature stabilized and pressure regulated room. The magnet is installed on vibrationally damped feet. (b) Contour plot of the magnetic field values on different angles on a circle with 2.5$\,$mm radius along the axis of the bore tube. The spatial homogeneity over the 12$\,$cm central region of the magnet is about 25$\,$ppm. Moreover, in a volume of one cubic centimeter in the center of the Penning trap stack, the spatial homogeneity of the magnetic field was determined to be in the order of a few ppm.}
		\label{fig:Magnetic-field}
	 \end{figure*}

\begin{figure*}
	\resizebox{.85\textwidth}{!}{
		\includegraphics{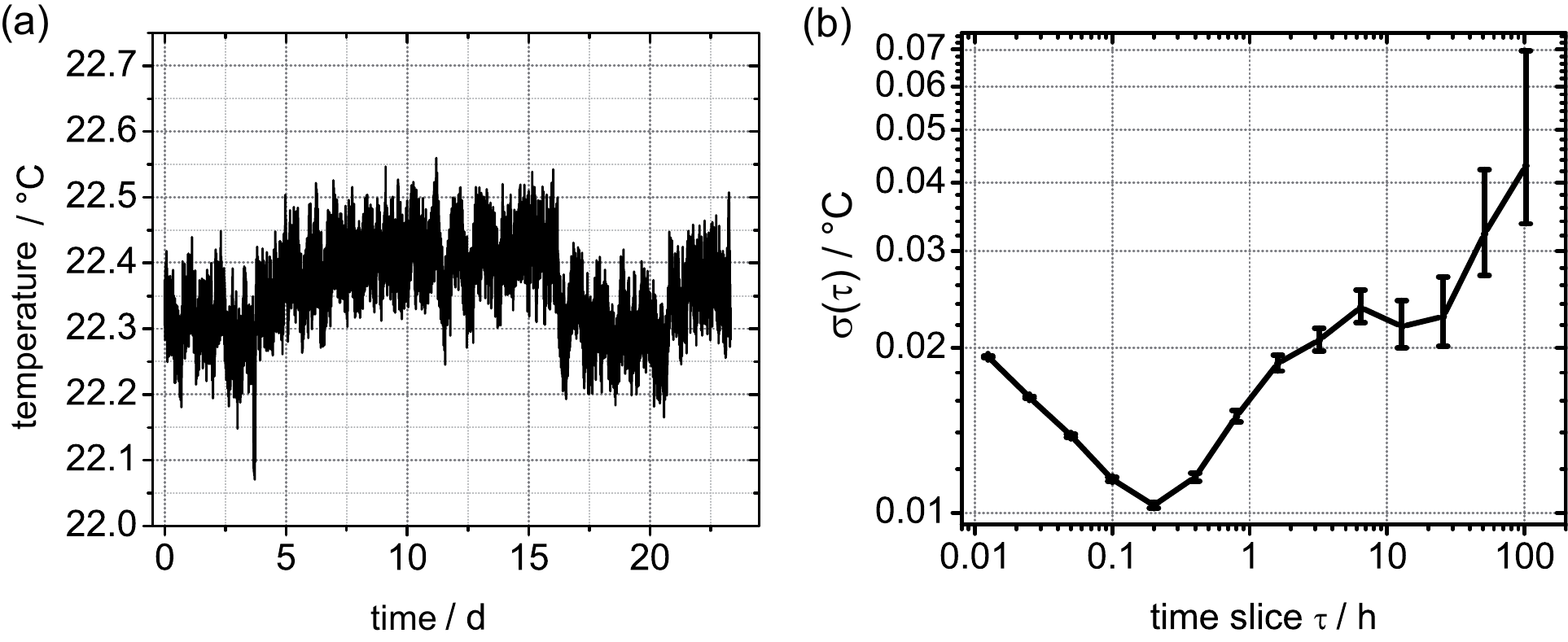}}
	\caption{Temperature behavior over approximately three weeks in a closed volume around the \textsc{Pentatrap} magnet (a) and the resulting overlapping Allan deviation (b) calculated with AlaVAR 5.2 \cite{AlaVAR52}. }
	\label{fig:PENTATRAPRaumbed-Temperatur}
\end{figure*}

\begin{figure*}
	\resizebox{.8\textwidth}{!}{
		\includegraphics{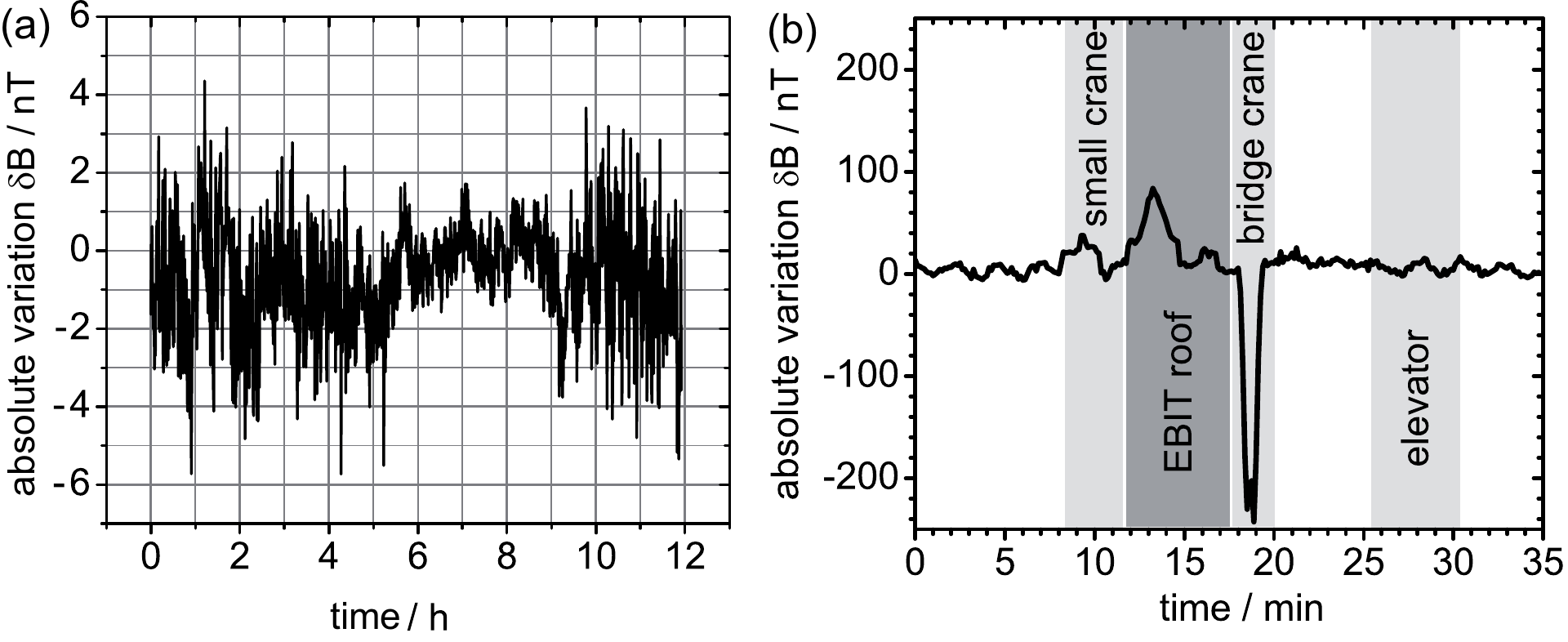}}
	\caption{(a) Vertical earth magnetic field at night. (b) Examination of environmental influences, such as the movement of several cranes, the roof of the EBIT lab, and the elevator in the building during a standard working day. In both cases, the field was measured with a flux-gate magnetometer. The Heidelberg-EBIT situated above the \textsc{Pentatrap} experiment was not in operation and the superconducting magnet had not yet been installed.}
	\label{fig:PENTATRAPRaumbed-Magnetfeld}
\end{figure*}

\begin{figure*}
	\resizebox{.6\textwidth}{!}{
		\includegraphics{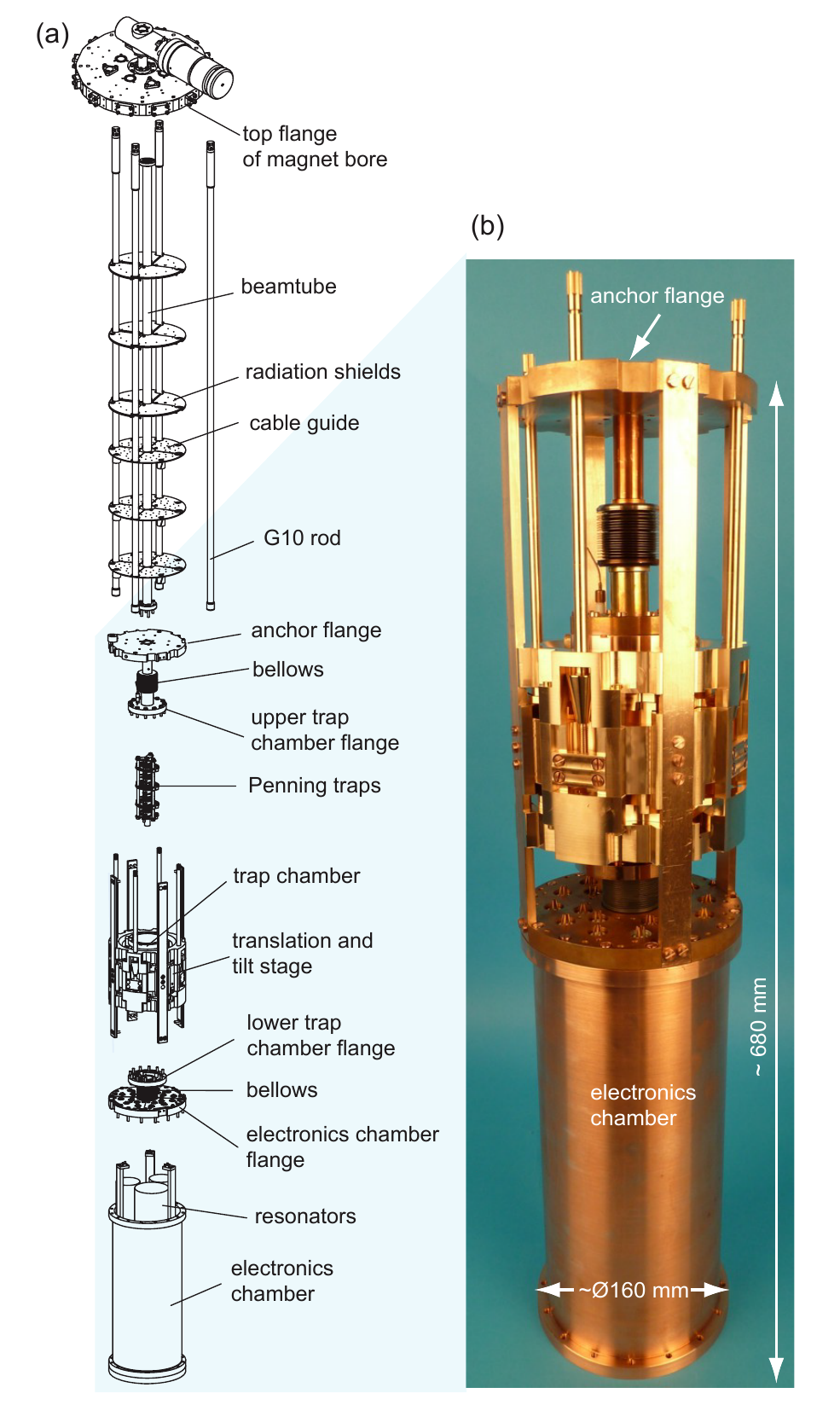}}
	\caption{(Color online) (a) Exploded assembly-drawing of the cryogenic part of \textsc{Pentatrap} with close to 1200 components in total. Main components are the trap chamber with the Penning trap stack inside, a translation and tilt stage outside, and the electronics chamber including resonators and amplifiers. An anchor flange will fix the vertical position of the system. To avoid heat transfer, several radiation shields as well as fiber glass rods (G10) are installed. (b) Photograph of a fraction of the cryogenic part.}
	\label{fig:Adjustment+Traps}
\end{figure*}

\begin{figure*}
\resizebox{.99\textwidth}{!}{
		\includegraphics{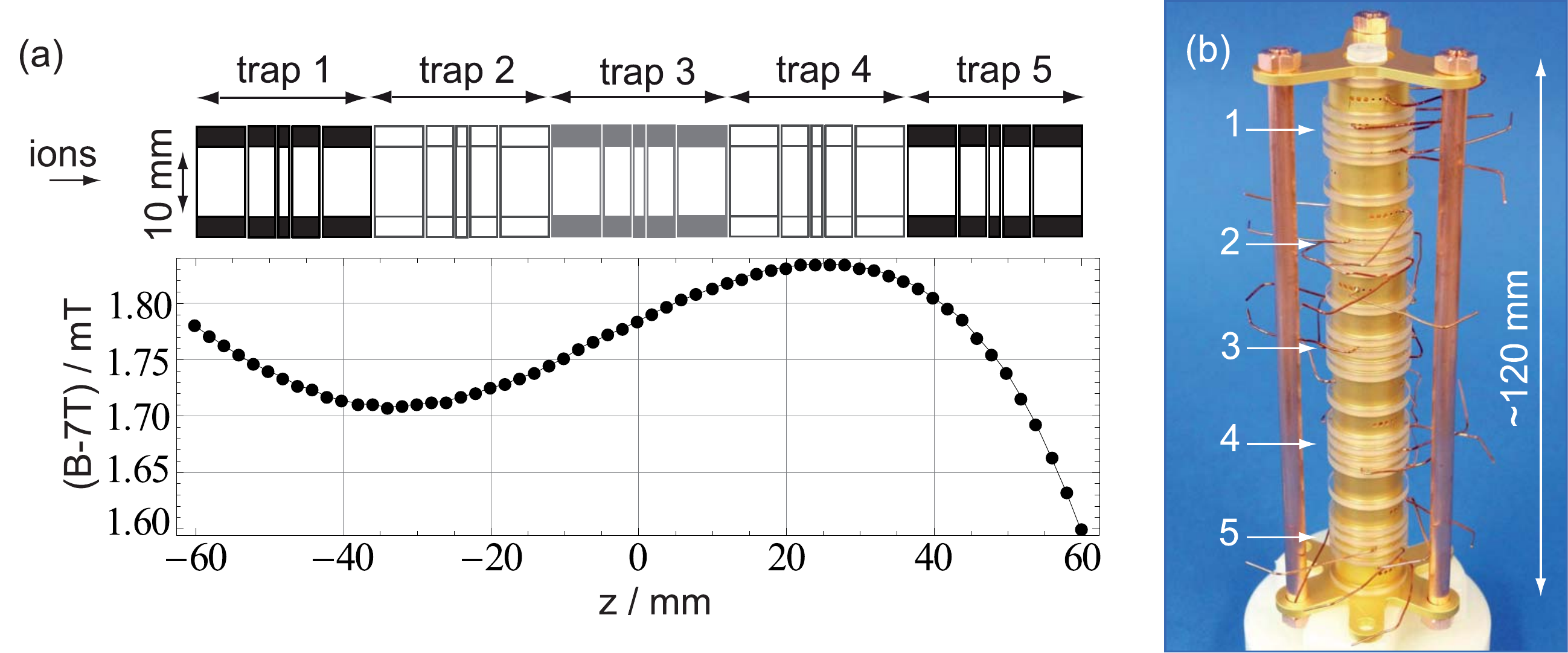}}
	\caption{(Color online)(a) Schematic of the five trap setup. The lower graph shows the magnetic field along the axis. The total length of the trap is constrained by the homogeneous region of the magnet (see also Fig.~\ref{fig:Magnetic-field}). For further details see text. (b) Photograph of the Penning trap assembly.}
	\label{fig:Traps}
\end{figure*}

The new high-precision mass spectrometer \textsc{Pentatrap} will be installed at the Max-Planck-Institut f\"ur Kernphysik in Heidelberg. An overview of the experimental setup is shown in Fig.~\ref{fig:PENTATRAPConcept}. 
In this section, the components of the mass spectrometer are discussed.
\\\\
\noindent
\textbf{Ion sources}\\
Well-proven sources for highly charged ions are electron beam ion traps (EBITs) \cite{Levine1988,Levine1989}. Such devices have demonstrated to deliver ions up to Cf$^{96+}$ \cite{Marrs1994a,Marrs1994b,Beiersdorfer1997}.
At \textsc{Pentatrap}, highly charged long-lived and stable ions will be provided by two EBITs: a small commercial room temperature Dresden-EBIT3 \cite{Ovsyannikov1999,Zschornack2010,Dreebit2011} (see Fig.~\ref{fig:EBIT3}(a)) with an additional Wien filter, and the Heidelberg-EBIT \cite{Crespo2000,Crespo2001,Crespo2004}.
\\
Dresden-EBIT3 is dedicated to the production of bare ions with nuclear charge numbers up to $Z\,$=$\,$30. 
For elements with higher nuclear charge number, complete ionization is not possible. Thus, only ions up to helium- or neon-like charge states can be produced for medium- and high-$Z$ elements, respectively. This ion source will be used for the commissioning of \textsc{Pentatrap} and for the investigation of its performance.
Moreover, this ion source can be used to produce ions for measurements in the field of neutrino physics. The production of osmium and rhenium ions with natural abundances with charge states up to about 50+ has been demonstrated (see Fig.~\ref{fig:EBIT3}(b)).
\\
For physics applications, which require extremely high charge states of long-lived and stable nuclides, e.g., bare or hydrogen-like lead or uranium, \textsc{Pentatrap} will obtain ions from the Heidelberg-EBIT.
Currently, the maximum electron beam energy of 100 keV allows a production of ions such as helium-like mercury Hg$^{78+} $\cite{Martinez2005} or hydrogenic barium Ba$^{55+}$, and stripping all electrons from elements up to krypton Kr$^{36+}$. For the extracted highly charged ion beam from the Heidelberg-EBIT, a beam emittance of 3$\cdot\pi\cdot$mm$\cdot$mrad has been demonstrated, which meets the requirements for the transport through the \textsc{Pentatrap} beamline and capture in the Penning traps.
In Tab.~\ref{tab:ChargeBreedingperformanceEBIT}, the calculated charge breeding performance of the Heidelberg-EBIT for some example ions at fixed electron beam energies is listed.
\begin{table}
\caption{Calculated charge breeding performance of the Heidelberg-EBIT for different electron beam energies $E_\mathrm{e}$. Parameters used for this calculation are an electron beam current of 500$\,$mA and an electron/ion overlap of 0.1. Here, $t_{\mathrm{ion}}$ and $C_{\mathrm{trap}}^{40\%}$ are the ionization time and the trap capacity at 40\,\% compensation of the ions. In general, the trap capacity represents the number of positive charges in the trap.}
\label{tab:ChargeBreedingperformanceEBIT}  
\begin{tabular}{lllll}
\hline\noalign{\smallskip}
ion&$E_\mathrm{e}$&ions in&$t_{\mathrm{ion}}$&$C_{\mathrm{trap}}^{40\%}$\\
	&(keV)&charge state&(s)&\\
\noalign{\smallskip}\hline\noalign{\smallskip}
				Ne$^{10+}$&5&60\,\%&0.016&7.2$\cdot$10$^7$\\
				Ar$^{18+}$&20&60\,\%&0.19&2.0$\cdot$10$^7$\\
				Kr$^{34+}$&60&60\,\%&0.54&6.1$\cdot$10$^6$\\
				Xe$^{52+}$&120&60\,\%&3.0&2.8$\cdot$10$^6$\\
				Pb$^{54+}$&160&10\,\%&0.27&3.9$\cdot$10$^5$\\
				Pb$^{72+}$&160&10\,\%&1.7&2.9$\cdot$10$^5$\\
				Pb$^{80+}$&160&10\,\%&7.4&2.6$\cdot$10$^5$\\
				Pb$^{81+}$&160&1\,\%&9.5&2.6$\cdot$10$^4$\\
				Pb$^{82+}$&160&0.1\,\%&17&2.6$\cdot$10$^3$\\
\noalign{\smallskip}\hline
\end{tabular}
\end{table}
In addition, \mbox{Fig.~\ref{fig:EBIT}} shows the calculated charge state production of iron ions as a function of the charge breeding time in the Heidelberg-EBIT.
An upgrade of this device, planned for the near future, will enable the production of bare uranium.
\\
Future prospects include plans to move the \textsc{Pentatrap} mass spectrometer to the \textsc{Hitrap} facility \cite{Kluge2008} at the Helmholtzzentrum f\"ur Schwerionenforschung GSI, where access to low-emittance beams of highly charged short-lived ions up to bare uranium with charge-specific kinetic energies of only a few kV$\cdot$q will be provided.
\\\\
\noindent
\textbf{Transfer beamline and beam diagnostics}\\
An approximately 2 meter long ion beamline will interface the Dresden-EBIT3 ion source with the Penning traps (see Fig.~\ref{fig:PENTATRAPConcept}). It consists of a series of electrostatic einzel lenses, two pulsed drift tubes, drift regions and two diagnostics stations. The einzel lenses allow an efficient transport of the ions towards the magnet and match the ion beam emittance to the acceptance of the magnetic field. Two pulsed drift tubes are used for a reduction of the kinetic energy of the ions from a few kV$\cdot$q down to few V$\cdot$q to ensure, together with the lens system, an efficient injection of the ions into the magnetic field, and to enable a capture of the ions in the traps. The ion transport from the ion sources to the magnet, the injection into the magnetic field, and the capture of the ions into the traps have been simulated with the software package \textsc{Simion}$^{TM}$ \cite{Manura2007} in order to optimize the voltage settings for an efficient transport. For an ion beam emittance of 3$\cdot\pi\cdot$mm$\cdot$mrad and a kinetic energy of 7 kV$\cdot$q, the efficiency of the simulated ion transport and injection into the magnetic field exceeds 95$\%$. The efficiency of the ion capture in the traps is a few ten percent for an ion bunch of a few hundred nanoseconds length.
\\
In order to monitor the ion beam transport, two movable diagnostic stations will be placed in the focal planes of the ion beamline. The first diagnostic station will be installed directly behind the Dresden-EBIT3 for a measurement and optimization of the ion beam parameters. The second diagnostic station will be mounted close to the entrance region of the magnet for monitoring the ion beam position and emittance right before the injection into the superconducting magnet. Figure \ref{fig:DiagnoseSystem}
shows the beam diagnostic station used at \textsc{Pentatrap}.
Each station is equipped with a Faraday cup (FC) to measure the ion beam current and a two inch multi-channel plate (MCP) detector with phosphor screen (PH) to monitor the size and position of the ion beam.
\\\\
\textbf{Superconducting magnet system}\\
\textsc{Pentatrap} utilizes a commercial 7 Tesla, actively\linebreak shielded (shielding factor $>$100) superconducting magnet with a vertical bore and an inner diameter of 160$\,$mm (Fig.~\ref{fig:Magnetic-field}(a)).
In the cold-bore ``bucket-type'' cryostat, the same helium reservoir is used to cool the magnet's superconducting coils as well as the Penning trap assembly. The spatial homogeneity of the magnetic field $\Delta B/B$ is in the order of a few ppm in the central 1$\,$cm$^3$ volume. The ions' oscillation amplitudes will not exceed a few ten $\mu$m \cite{Sturm2011a}. In the total trap volume, defined by a cylinder of 2.5$\,$mm radius and 120$\,$mm length along the axis of the bore tube, the spatial homogeneity amounts to about 25$\,$ppm. The measured magnetic field profile is shown as contour plot in Fig.~\ref{fig:Magnetic-field}(b).
The magnetic field gradient in the region of the traps will be flattened by compensation coils wound around the trap chamber.
\\
A stabilization system is planned to improve the overall temporal stability of the magnetic field $\Delta B/B\cdot 1/\Delta T$ to at least a few ten ppt per hour. It has already been shown that such stabilities can be reached \cite{VanDyck1999}.
In the following, the main characteristics of the stabilization system are presented.
\\
A main feature is the stabilization of the pressure and level in the liquid helium reservoir of the superconducting magnet. The helium level control will be based on the stabilization of the resonance frequency of a tuned circuit which is placed partially in liquid helium. The pressure stabilization is based on a temperature-stabilized gas vessel that serves as a pressure reference. Attainable values for the level and the pressure stabilities are estimated to be in the 0.1$\,$mm and $\mu$bar range, respectively.
\\
Moreover, in order to reduce external influences on the magnetic field, the magnet is positioned in a vibration-damped, temperature-stabilized and pressure-regulated room. The peak-to-peak variation of the temperature over one day is specified below 0.1$\,$K. Figure \ref{fig:PENTATRAPRaumbed-Temperatur} shows the measured temperature over about three weeks as well as the overlapping Allan deviation \cite{Allan1966} calculated from these values.
For a time slice of one day, the Allan deviation is about 0.02$\,$K. However, the measurement was carried out with almost no electrical power load in the lab.
The magnet is equipped with three anti-vibration pneumatic pads and rests on a vibration-damped, 70$\,$cm thick cushion made of concrete. The vibration amplitude of the cushion does not exceed 1 $\mu$m. An aluminum housing around the magnet will be installed for shielding the magnet against stray electric and high-frequency magnetic fields.
\\
Furthermore, the vertical component of large-scale, low-frequency magnetic fields will be compensated via a feedback control system based on a flux-gate magnetometer and a pair of Helmholtz coils positioned around the magnet \cite{VanDyck1999}. To correct these fluctuations, a current is generated which is proportional to the current required to null the variations in the magnetic field at the location of the flux-gate. This current is scaled and applied to the Helmholtz coil pair. An overall shield factor, which is the product of the compensation factor of the Helmholtz coils and the self-shielding factor of the magnet is estimated to exceed 1000.
Figure \ref{fig:PENTATRAPRaumbed-Magnetfeld}(a) shows the magnetic field in the magnet room measured with the flux-gate magnetometer at night, when there is less environmental interference than during daytime.
The effect of disturbing factors on the magnetic field is shown in Fig.~\ref{fig:PENTATRAPRaumbed-Magnetfeld}(b). The biggest disturbance is caused by the movement of the heavy duty bridge crane inside the accelerator building above the Penning trap laboratory. 
\\\\
\textbf{Cryogenic assembly}\\
The part of the system which is inserted into the bore of the magnet and thereby cooled to a temperature of 4$\,$K is called the cryogenic assembly (see Fig.~\ref{fig:PENTATRAPConcept} and Fig.~\ref{fig:Adjustment+Traps}).
It consists of two joined copper vacuum chambers for the trap and the electronics, which are connected to the warm beamline at the top of the magnet with a one meter long DN16 stainless steel tube. The trap chamber, containing a stack of five cylindrical traps, is mechanically connected to the electronics chamber and the stainless steel tube with two flexible titanium bellows. In combination with a translation and tilt stage, this enables an independent adjustment of the angular ($\pm$1$^\circ$) and horizontal ($\pm$2.2$\,$mm) positions of the trap´s with respect to the axis of the magnetic field. 
Despite the open system, an ultra-high vacuum of better than 10$^{-13}\,$mbar in both chambers is achieved through cryo-pumping at the 4$\,$K surface and by using charcoal absorbers in both chambers. Since the electronics chamber houses ``in-vacuum" cryogenic resonators and amplifiers (see Chap.~4), a direct coupling of the resonators to the trap electrodes is provided.
\\
Materials were chosen thoroughly with respect to appropriate mechanical properties, such as matching thermal expansion coefficients and low magnetic susceptibilities. The trap chamber, as well as the electronics chamber are made of oxygen-free, high thermal conductivity (OFHC) copper with a purity greater than 99.99$\%$. The bellows are made of titanium grade 2, which remains flexible at 4$\,$K. The components of the translation and tilt device (Fig.~\ref{fig:Adjustment+Traps}) are made of phosphorous bronze (CuSn8). Fiber glass (G10) tubes are used to actuate the translation and tilt device. Special, electrical non-magnetic copper feed\-throughs without any nickel alloy, which is the standard brazing material, are used. For thermal insulation, highly polished aluminum radiation shields are installed.
\\\\
\textbf{Penning traps}\\
A stack of five identical, cylindrical, orthogonalized and compensated five-electrode traps comprises the heart of \textsc{Pentatrap} (see Tab.~\ref{tab:trapprop} and Fig.~\ref{fig:Traps}). The compensation of the traps provides a harmonic electric potential in the traps' central regions by tuning the applied voltages. The orthogonalization of the traps makes the axial oscillation frequency of the particle less sensitive to the correction voltage.
In general, the electrostatic potential of a Penning trap can be expressed throughout a Taylor expansion \cite{Verdu2008},
and the influence of the electric field coefficients on the ion modes can be determined as done for the cylindrical trap in \cite{Gabrielse1984}.
For the analysis of the electrostatic potential and the elaborate design studies of the five traps of \textsc{Pentatrap}, see the following article in this issue \cite{Roux2011}, in which image charge effects, mutual influence of the trapping potential of the adjacent traps, ion-ion interaction between two ions stored in different traps and the influence of machining tolerances on the performance of the traps are considered.
\\ 
In order to keep influences on the magnetic bottle term small, low magnetic susceptibility materials are used for the traps. Thus, the electrodes are made of $>$99.999\,$\%$ OFHC copper, and the insulators between them of sapphire.
Moreover, patch potentials on the electrode surfaces have to be avoided, since they influence the electric storage potential seen by the ion. Therefore, the trap electrodes are galvanically gold-plated with a 20$\,\mu$m thick layer in a pure gold bath without any organic or inorganic additives that are commonly used to enhance shine. The traps are fabricated with a tolerance of about $\pm$2$\,\mu$m, and galvanically processed with a tolerance of better than $\pm$3$\,\mu$m, leading to a total tolerance of about 5$\,\mu$m.\\
Measurements of the cyclotron frequencies of ions of interest will be performed in the three inner traps, labeled 2-4 in Fig.~\ref{fig:Traps}, whereas the outer traps 1 and 5 will serve for ion storage, and in some cases, for monitoring magnetic field fluctuations or as a reference for the voltage source. The measurement procedure and individual functions of the particular traps are discussed in detail in Sec.~\ref{measurementprocedure}.
\begin{table}
\caption{Main trap parameters of a single trap. The dimensions of the Penning trap electrodes as well as the dependence of the axial frequency on the ratio $\mathrm{TR}=U_{\mathrm{C}}/U_{\mathrm{0}}$ (tuning ratio) of the correction $U_{\mathrm{C}}$ and the ring voltage $U_{\mathrm{0}}$ are shown. In the case of a vanishing fourth order expansion coefficient of the electric field, the tuning ratio is 0.881. A measure of the influence of the correction voltage on the axial frequency compared to the influence of the ring voltage is the term $d_2/c_2$, with $d_2\,$=$\,\partial c_2/\partial \mathrm{TR}$. For more details, especially the uncertainties of the last two values, see \cite{Roux2011}.}
\label{tab:trapprop}      
\begin{tabular}{lll}
\hline\noalign{\smallskip}
Parameter&&Value\\
\noalign{\smallskip}\hline\noalign{\smallskip}
overall length & & 24$\,$mm\\
& endcap electrode& 7.040$\,$mm \\
length of& correction electrode & 3.932$\,$mm \\
& ring electrode & 1.457$\,$mm \\	
inner radius&&5$\,$mm\\
\multicolumn{2}{l}{gaps between electrodes}&0.15$\,$mm\\
$|\Delta\nu_z/\Delta \mathrm{TR}|$&&$<\,$2.5$\,$Hz/mUnit\\
$d_2/c_2$&&$<\,$0.01\\
\noalign{\smallskip}\hline
\end{tabular}
\end{table}

\section{Ion detection}
\label{Iondetection}
\begin{figure*}[htb]
	\resizebox{.99\textwidth}{!}{
		\includegraphics{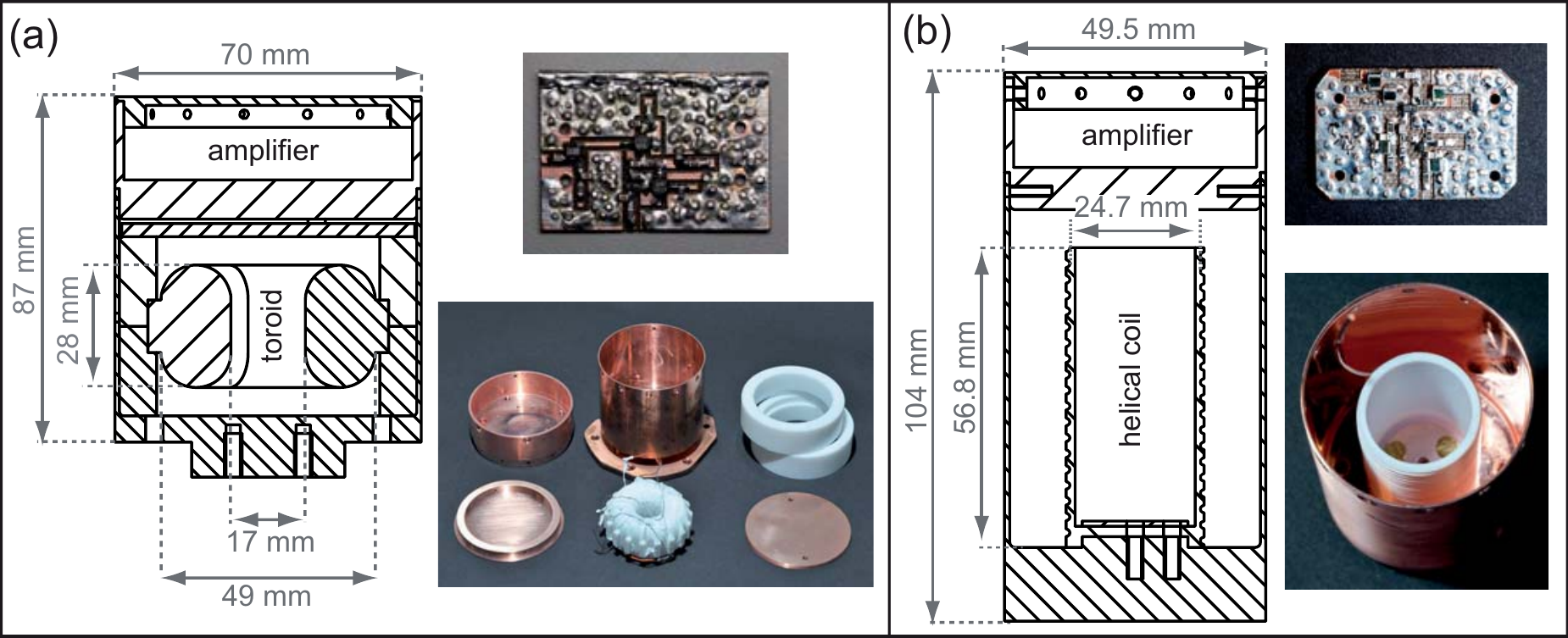}}
	\caption{(Color online)(a) Schematic and photograph of the axial detection system. The superconducting toroidal coil is made of a few hundred ($\approx\,$800) niobium-titanium windings around a teflon torus and is placed in a cylindrical housing made of high-purity OFHC-copper. The amplifier has a common-source input stage with high input impedance and extremely low equivalent input noise. The system will be operated at $\nu_z\approx\,$600$\,$kHz. (b) Schematic and photograph of the cyclotron detection system, which is designed for the detection of ions with $m/q$ values between about 4.0$\,$u/e and 4.5$\,$u/e corresponding to resonance frequencies $\nu_+$ between about 24$\,$MHz and 27$\,$MHz. The cyclotron detection coil has solenoidal geometry and copper windings.}
	\label{fig:ampcoil}
\end{figure*}

\begin{figure*}[htb]
	\resizebox{.99\textwidth}{!}{
		\includegraphics{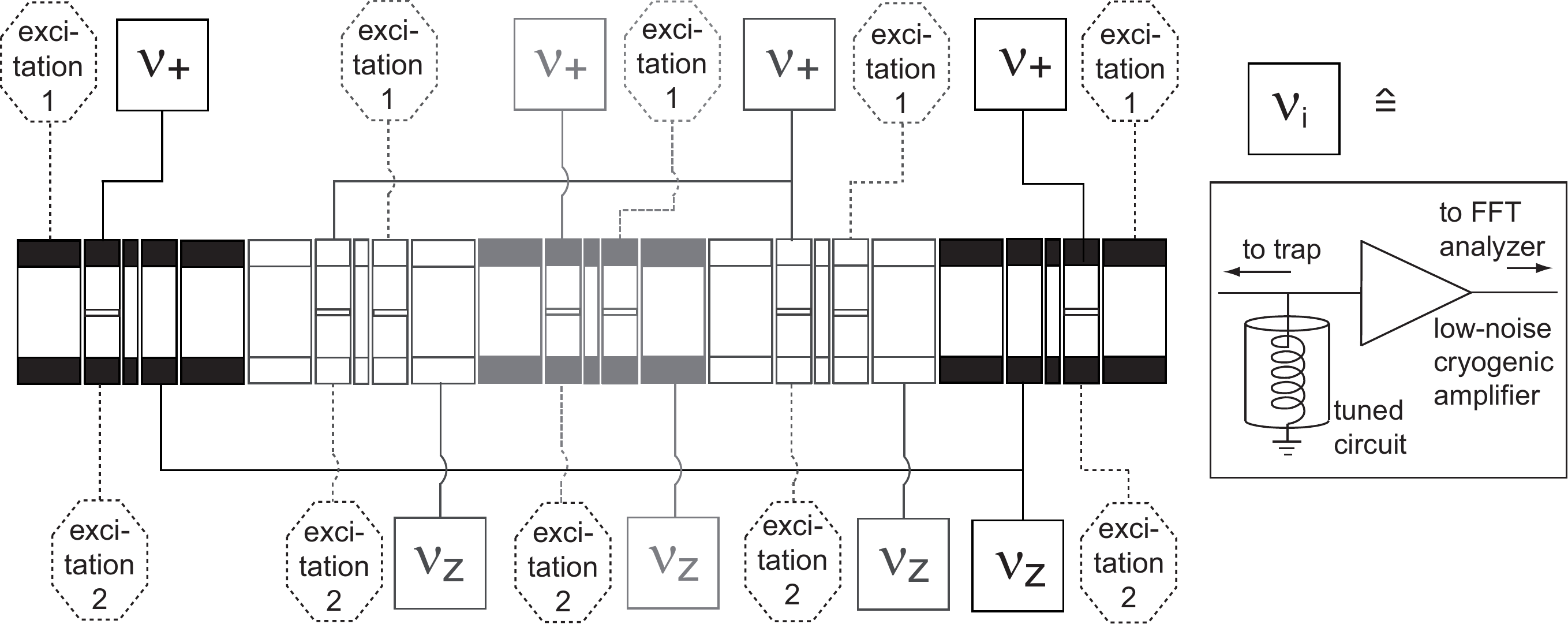}}
			\caption{Sketch of a proposed cryogenic detection and excitation system for \textsc{Pentatrap} consisting of four axial and four radial tank circuits with high-$Q$ inductors. Low-noise cryogenic amplifiers will complete the cryogenic detection system. Each trap will be connected to two excitation lines to apply dipolar as well as quadrupolar excitations. The position of the segmented electrodes is shown.}
	\label{fig:Detection-circuit}
\end{figure*}

The variation of the cyclotron frequency as a function of the variation of the ion's eigenfrequencies follows from Eq.~(\ref{eq:invariancetheorem}) and is given by
\begin{eqnarray}
\frac{\Delta\nu_c}{\nu_c}&=&\left(\frac{\nu_+}{\nu_c}\right)^2\cdot\frac{\Delta\nu_+}{\nu_+}+\left(\frac{\nu_z}{\nu_c}\right)^2\cdot\frac{\Delta\nu_z}{\nu_z}\nonumber\\&&+\left(\frac{\nu_-}{\nu_c}\right)^2\frac{\Delta\nu_-}{\nu_-}.
\label{relativeprecision}
\end{eqnarray}
For example, the eigenfrequencies of $^{187}$Os$^{45+}$ are \linebreak $\nu_+\,$=$\,$25.86$\,$MHz, $\nu_-\,$=$\,$6.96$\,$kHz and $\nu_z\,$=$\,$600$\,$kHz, corresponding to a magnetic field of 7$\,$T and a trapping voltage $U_0\,$=$\,-$20.45$\,$V. In order to measure $\nu_c$ with a precision of a few parts in 10$^{12}$, one has to measure $\nu_+$, $\nu_z$ and $\nu_-$ with a relative precision of about 10$^{-12}$, 10$^{-9}$ and 10$^{-5}$, respectively. 
Consequently, the main challenge is to achieve the necessary precision in the determination of $\nu_+$. In this section, the basic principles of the ion detection and the resulting experimental constraints are discussed.
\\
At \textsc{Pentatrap}, the non-destructive measurement of the eigenfrequencies of the stored ions is carried out with the image-current detection technique \cite{Wineland1975}. As a result of the ions' oscillation, image currents $I$ 
\begin{equation}
I=2\pi\frac{q}{D}\nu_i\rho_i
\label{imagecurrent}
\end{equation}
are induced in the trap electrodes, where $\nu_i$ is the respective eigenfrequency and $\rho_i$ the corresponding motional amplitude. $D$ is a characteristic length defined by the geometry of the trap. Typically the induced currents $I$ are in the fA range. These tiny currents can be detected with highly sensitive detection systems, which are connected to the trap electrodes. 
Such a detection system, technically described in \cite{Jefferts1993}, employs a tank circuit that consists of an inductor $L$ and the system's total capacitance $C$. The total capacitance $C=C_\mathrm{t}+C_\mathrm{c}+C_\mathrm{p}$ is given by the trap capacitance $C_\mathrm{t}$, the capacitance of the coil $C_\mathrm{c}$ and other parasitic capacitances $C_\mathrm{p}$. At the tank circuit's resonance frequency $2\pi\nu_0\,$=($LC$)$^{-1/2}$, the effective resistance is $R_{\mathrm{res}}\,$=$\,2\pi\nu_0 Q L$, where $Q$ is the quality factor of the tank circuit. Tuning the resonance frequency of the detector to the eigenfrequency of the particle, the tiny image current $I$ causes a voltage drop $V\,$=$\,R_{\mathrm{res}} I$, which is amplified by an ultra low-noise amplifier and analyzed with a FFT spectrum analyzer.
\\
In order to work at high signal-to-noise ratios, the whole trap system is operated at cryogenic temperatures (4$\,$K). In this environment, superconducting alloys (NbTi) can be used for the inductors, which leads to high quality factors \cite{Ulmer2009}. Furthermore, at 4$\,$K the equivalent input noise density $e_\mathrm{n}$ of the amplifiers is reduced, e.g., $e_\mathrm{n}$ amounts to approximately 700$\,$pV/Hz$^{1/2}$ for the axial amplifier at $\nu_0\,$=$\,$600$\,$kHz.
The field-effect-transistors (FETs) used for the amplifiers are based on GaAs. Due to the small band-gap of this semiconductor, charge carriers do not freeze out completely and suitable performance under cryogenic conditions is assured. The common-source amplifiers have a high input resistance better than 10$\,$M$\mathrm{\Omega}$ and provide an amplification higher than 10$\,$dB at cryogenic temperatures. In Tab.~\ref{tab:axialandcyclotron},
\begin{table}
\caption{Measured values of the inductance $L$, capacitance $C_\mathrm{c}$, number of turns $N$, and unloaded quality factor $Q^*$ of the detection coils designed for the \textsc{Pentatrap} experiment. Connected to the Penning trap, resonance frequencies of about 600$\,$kHz for the axial, and of about 27$\,$MHz for the modified cyclotron frequency detection will be realized.}
	\label{tab:axialandcyclotron}    
\begin{tabular}{lllll}
\hline\noalign{\smallskip}
Coil&$L$ ($\mu$H)& $C_\mathrm{c}$ (pF)& $N$ & $Q^*$\\
\noalign{\smallskip}\hline\noalign{\smallskip}
toroidal axial&$\approx\,$3400&$\approx\,$7&$\approx\,$800&$>\,$75000\\
helical cyclotron&$\approx\,$1.9&$\approx\,$3.2&13&$>\,$3550\\
\end{tabular}
\end{table}
typical parameters of the axial and cyclotron detection coils are listed. Fig.~\ref{fig:ampcoil}
shows the overall detection system as schematic and photograph.
\\
When coupled to the detection system, the particle cools resistively to the ambient temperature of the detection system \cite{Wineland1975} with a cooling time constant
\begin{equation}
\tau=\frac{m}{q^2}\frac{D^2}{R_{\mathrm{res}}}.
\label{coolingconstant}
\end{equation}
Depending on the electrode chosen for the axial signal pickup and for typical values of the loaded $Q$, the estimated cooling constant at \textsc{Pentatrap} ranges from a few seconds for low charged ions down to a few milliseconds for highly charged ions.
The cooling time constant of the reduced cyclotron motion is estimated to range from seconds to a few minutes for highly and low charged ions, respectively. Thus, higher charge states are preferable because the measurement cycle can be kept shorter and the ions are confined to a smaller volume.
When the ions are in thermal equilibrium with the detection circuit, which is in direct contact with the liquid helium at 4$\,$K, they short out the thermal noise of the tuned circuit, and a dip occurs in the FFT spectrum with a minimum at the eigenfrequency of the particle \cite{Wineland1975}. Thereby, the respective eigenfrequency $\nu_i$ of the trapped ion can be directly measured by determining the frequency of the minimum in the noise spectrum of the detection system. In a frequency measurement, which is carried out detecting the noise dip, the trapped particle is at low temperatures, which corresponds to low amplitudes. Thus, the ion is less affected by electric and magnetic field errors.
\\
Due to the very narrow and commonly not resolvable noise dip of the modified cyclotron frequency $\nu_+$, the noise-dip detection technique is applied only to measure the axial frequency $\nu_z$ of the particle.
The radial frequencies $\nu_+$ and $\nu_-$ are measured indirectly via resonant sideband coupling \cite{Cornell1990}. 
Applying a radiofrequency drive signal with frequency $\nu_{\mathrm{rf}}\,$=$\,\nu_\pm\mp\nu_z$ to the trap, a double-dip structure occurs in the thermal noise spectrum of the detector. By determining the frequencies of both dips, the modified cyclotron frequency $\nu_+$ and the magnetron frequency $\nu_-$, respectively, can be calculated as described in \cite{Cornell1990}.
\\
Another possible technique to measure the modified cyclotron frequency $\nu_+$ uses phase-sensitive detection. Compared to the double-dip detection technique, it is faster \cite{Stahl2005} and therefore less sensitive to magnetic and electric field fluctuations. Phase fluctuations are a limiting factor, which can only be minimized through longer measurement times or by increasing the signal-to-noise ratio. The latter can be achieved by an excitation of the particle's motional amplitudes. In this context, the use of highly charged ions is advantageous since lower amplitudes are needed to obtain a sufficient signal, and therefore the ion experiences fewer anharmonicities of the storage fields. More details are given in \cite{Roux2011}.
\\
A well-proven phase-sensitive method to measure the modified cyclotron frequency $\nu_+$ with low uncertainty down to a level of 10$^{-11}$ is the so-called Pulse-aNd-Phase (PNP) technique \cite{Cornell1989}. 
In this technique, the modified cyclotron motion of the previously cooled ion is first excited to a
fixed amplitude and starting phase. After some well-defined phase evolution time, a coupling pulse is applied that swaps the phase of the cyclotron and axial modes. Thus, the modified cyclotron phase can be determined by measuring the axial phase after different phase evolution times. The knowledge of the modified cyclotron phase and the phase evolution time leads to a determination of the modified cyclotron frequency.
\\
For the ultra high-precision experiments with \textsc{Pentatrap}, it is planned to apply a novel phase-sensitive technique introduced by S.~Sturm et al.~\cite{Sturm2011a} for the measurements of $\nu_+$. Similar to the well-known PNP technique this Advanced-Pulse-aNd-Phase (APNP) technique allows the direct extraction of the phase information of the cyclotron motion from the digitized axial signal. 
In the APNP method, in contrast to the PNP method, cyclotron-to-axial-coupling is applied at the sum frequency, producing parametric amplification of the resulting axial motion. This allows cyclotron measurements with substantial smaller initial cyclotron amplitudes, in principle independent of the detector performance, hence reducing energy-dependent systematic shifts. This allows a measurement of $\nu_+$ with higher precision in a shorter time.
\\
The main experimental features to achieve the required precision are a 4$\,$K environment, a stable magnet system (see Sec.~\ref{Expsetup}) and fast measurement cycles (see Sec.~\ref{measurementprocedure}) to reduce the effect of magnetic noise, as well as a stable voltage source to achieve $\Delta\nu_z/\nu_z\,\leq\,$10$^{-9}$. Therefore, even when $\nu_z$ is taken as the average over many measurements, the ring voltage must be stable on a level of a few parts in 10$^8$ over a few minutes.
There are no commercial voltage sources available which fulfill these requirements. To this end, the electronics workshop at MPIK in collaboration with the Physikalisch-Technische Bundesanstalt Braunschweig (PTB), Germany, is currently developing a highly stable voltage source covering a voltage range between 0 and $-$100$\,$V.
Design parameters are a voltage stability of $<\,$4$\cdot$10$^{-8}\,$/(10$\,$min), a resolution of $<\,$10$\,\mu$V and a temperature stability of $<\,$4$\cdot$10$^{-7}\,$/K.
\\
\textsc{Pentatrap} aims to address a broad spectrum of $q/m$-ratios. Therefore, axial tank circuits are the main detection circuits in the inner traps 2, 3 and 4 (see Fig.~\ref{fig:Traps}). Different charge-to-mass ratios can then be brought in resonance with the circuits by simply changing the trapping voltage (see Eq.~(\ref{eq:axialfrequency})) and the radial frequencies $\nu_\pm$ will be measured via their coupling to the axial tank circuit as previously discussed. 
For test purposes the inner traps will be equipped with cyclotron tank circuits whose resonance frequencies can be tuned to a specific $q/m$-ratio. In the outer traps 1 and 5 (see Fig.~\ref{fig:Traps}) the fluctuations of the magnetic field will be monitored by a direct measurement of $\nu_+$ of, e.g., He$^+$ ions. Here, the main detection circuit is the cyclotron tank circuit, and the axial tank circuit will serve for test purposes, measurements of fluctuations of $U_0$ as well as for the ion preparation. 
Figure \ref{fig:Detection-circuit} schematically  shows a planned configuration of the detection circuits. For manipulating the eigenmotions of the stored ions, an axial and radial dipolar excitation will be possible in each trap. Moreover, the design allows for a mode coupling via a quadrupolar excitation.
The DC-voltages for the Penning trap and amplifier supply will be filtered at room temperature as well as at 4$\,$K. The ion signal has to be amplified again at room temperature. Therefore, four cyclotron room temperature amplifiers with subsequent mixers (\textit{ZFL-500LN} and \textit{ZAD6+} from \textit{Mini-Circuits}), as well as four axial room temperature amplifier boards with included down converters (\textit{AF-DC-b} from \textit{Stahl Electronics}) are situated on the magnet's top flange.

\section{Measurement procedure}
\label{measurementprocedure}
\begin{figure*}[htb]
\resizebox{.85\textwidth}{!}{
		\includegraphics{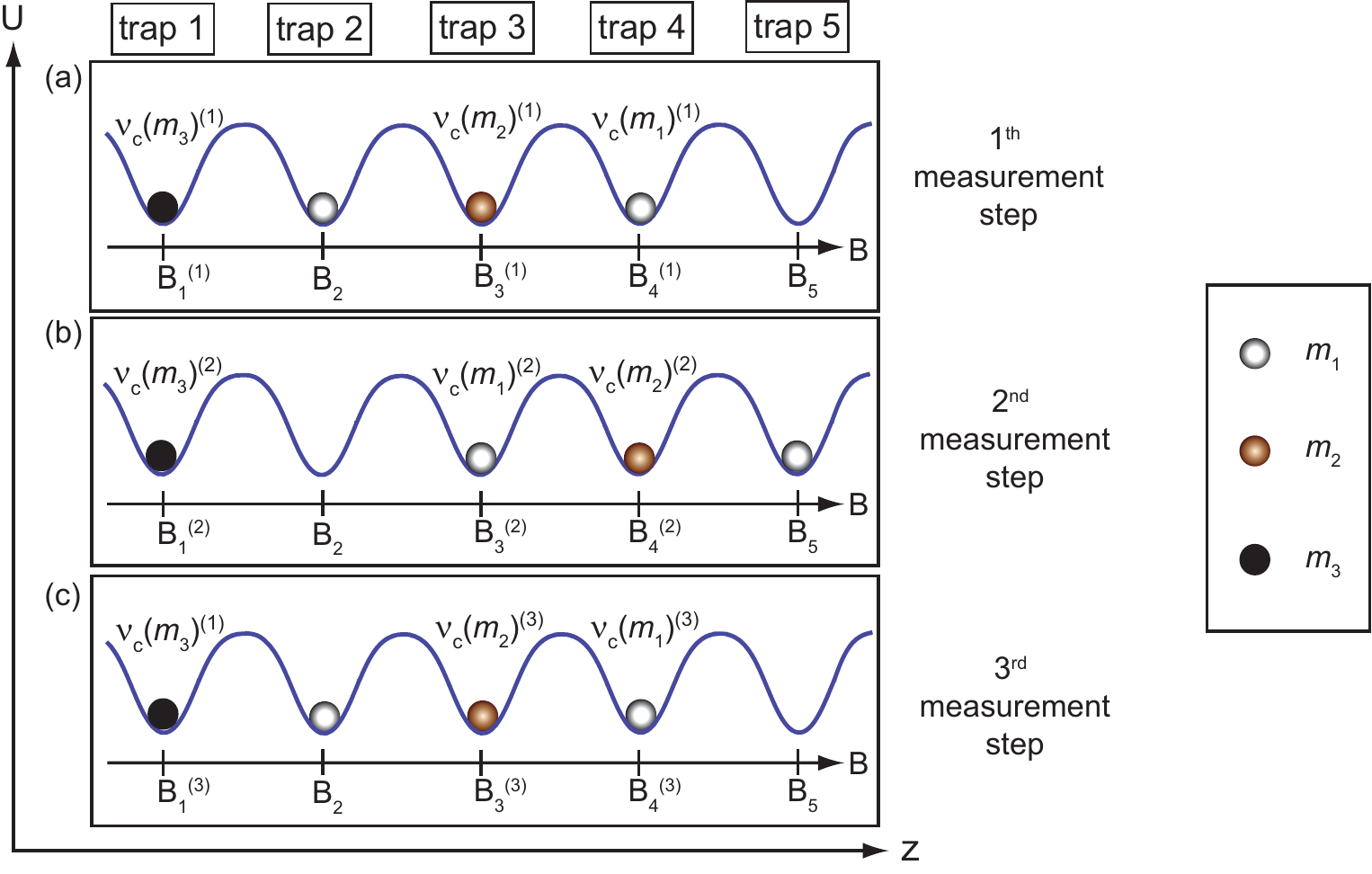}}	
		\caption{(Color online) Proposed measurement scheme of first choice. The potential wells are shown as a solid blue line. Two ions with mass $m_1$ are shown as grey balls, the ion with mass $m_2$ as a brown ball and the permanently stored ion with mass $m_3$ as a black ball. (a) In step 1, three ions with mass $m_1$, $m_2$ and $m_1$ are loaded into traps 2, 3 and 4, respectively. The cyclotron frequencies are measured in trap 3 and 4. (b) In step 2 the ions are moved one trap further to the right and again the cyclotron frequencies of the ions are measured in trap 3 and 4. (c) The ions are moved back to the initial position and the same measurement as in (a) is performed again. Trap 1 can be used for a longterm monitoring of the cyclotron frequency of the ion with mass $m_3$. These cycles will be repeated several times.}
	\label{fig:measpreocedure}
\end{figure*}

\textsc{Pentatrap} is the first Penning trap experiment to use in one setup a stack of five Penning traps for high-precision mass measurements. 
Since the effect of magnetic field fluctuations and drifts generally increases with time, it is indispensable that both the measurement of the frequency ratio and the particle preparation are done fast. An application of an FT-ICR measurement principle where two ions are stored simultaneously in one trap and drifts consequently cancel out since they affect the ions in the same way, is not applicable in our case. The perturbation due to the Coulomb interaction of the highly charged ions would increase the systematics and would make a high-precision mass measurement impossible \cite{Roux2011}. A solution is the use of a set of Penning traps, assuming that the temporal variation in the magnetic field is largely common to the different traps.
Moreover, with a set of five Penning traps a variety of measurement schemes is possible to determine mass ratios. 
\\
A measurement scheme of first choice implies the usage of traps 3 and 4 (see Fig.~\ref{fig:Traps}) to measure the cyclotron frequencies of two ionic species simultaneously. The measurement sequence is carried out as follows (see Fig.~\ref{fig:measpreocedure}):
First, ions with the same charge state and with masses $m_1$, $m_2$ and $m_1$ are loaded into traps 2, 3 and 4, respectively (Fig.~\ref{fig:measpreocedure}(a)).
The cyclotron frequencies of the ions $\nu_c(m_2)^{(1)}$ and $\nu_c(m_1)^{(1)}$ in traps 3 and 4 are measured simultaneously, and the ratio of their frequencies $R^{(1)}\,$=$\,\nu_c(m_2)^{(1)} /\nu_c(m_1)^{(1)}$ is determined according to Eq.~(\ref{eq:cyclotronfrequency}) to be $R^{(1)}\,$=$m_1/m_2\cdot B_3^{(1)}/B_4^{(1)}$, where $B_3^{(1)}$ and $B_4^{(1)}$ are the magnetic field strengths in trap 3 and 4 averaged over this measurement. After the measurement is completed, the ions are adiabatically moved into the adjacent traps (Fig.~\ref{fig:measpreocedure}(b)), so that trap 2 becomes empty, traps 3 and 4 are loaded with the ions with mass $m_1$ and $m_2$. The ratio of the cyclotron frequencies is measured again, $R^{(2)}\,$=$\,\nu_c(m_2)^{(2)} /\nu_c(m_1)^{(2)}\,$= $m_1/m_2\cdot B_4^{(2)}/B_3^{(2)}$, where $\,B_4^{(2)}$ and $B_3^{(2)}$ are the magnetic field strengths in trap 4 and 3, averaged over the second measurement period. The square root of the product $R^{(1)}$ and $R^{(2)}$ yields the $m_1$/$m_2$-ratio, assuming the ratio of the magnetic field strengths is time-independent. Thus, unavoidable temporal magnetic field fluctuations common to traps 3 and 4 do not affect the mass ratio. In this measurement scheme, trap 2 and 5 are used to park ions, which are temporarily not measured. In this approach, trap 1 can be used for a long-term monitoring of magnetic field fluctuations or as a reference for the voltage source. To this end, the cyclotron frequency of a permanently stored ion with mass $m_3$ can be monitored during the measurement cycle.
\\
The LabVIEW-based control and data taking system for the different measurement sequences is based on the GSI Control System (CS) framework \cite{Beck2003,Beck2004}.

\section{Conclusion and outlook}
\label{ConclusionOutlook}
A world-wide unique new Penning trap project for high-precision mass measurements on highly charged, stable, and long-lived ions up to uranium is presented, and the planned experimental setup is described. A combination of five traps will allow - for the first time - fast measurement cycles in which the ion exchange takes a few hundred milliseconds, in combination with a continuous observation of the magnetic field fluctuations during the whole measurement process. The benefits of externally produced highly charged ions will be combined with an advanced cryogenic trapping and non-destructive ion detection system.
\textsc{Pentatrap} aims for high-precision mass-ratio measurements of mass doublets with an uncertainty of a few parts in 10$^{12}$. In 2012, the complete facility will be assembled and extensive studies of its performance will be carried out.
\\
This work is supported by the Max Planck Society and by the Deutsche Forschungsgemeinschaft under contract BL 981/2-1. We thank D.~Beck, C.~Diehl, F.~Herfurth, M.~H\"ocker, J.~Ketter, H.-J.~Kluge, D.~Pinegar, W.~Quint, S.~Streubel and R.~Zirpel for stimulating discussions. Yu.~N. and Ch.~B. acknowledge the support from the Extreme Matter Institute (EMMI). S.~U. acknowledges support from the \textsc{Imprs-qd}. The excellent work by the MPIK construction office, and the mechanics and electronics workshop is highly acknowledged.


\begin{thebibliography}{}

 \bibitem{Blaum2006}K.~Blaum, Phys.~Rep.~\textbf{425,} (2006) 1.
 \bibitem{Franzke2008}B.~Franzke, M.~Geissel, G. M\"unzenberg, Mass Spectrom.~Rev.~\textbf{27,} (2008) 428.
 \bibitem{Blaum2010}K.~Blaum, Yu.~N.~Novikov, G.~Werth, Contemp.~Phys.~\textbf{51,} (2010) 149.
 \bibitem{Lunney2003}D.~Lunney, J.~M.~Pearson, C.~Thibault, Rev.~Mod.~Phys.~\textbf{75,} (2003) 75.
 \bibitem{Blaum2003}K.~Blaum, G.~Audi, D.~Beck, G.~Bollen, F.~Herfurth, A.~Kellerbauer, H.-J.~Kluge, E.~Sauvan, S.~Schwarz, Phys.~Rev.~Lett.~~\textbf{91,} (2003) 260801.
 \bibitem{Kankainen2010}A.~Kankainen, T.~Eronen, D.~Gorelov, J.~ Hakala, A.~Jokinen, V.~S.~Kolhinen, M.~Reponen, J.~Rissanen, A.~Saastamoinen, V.~Sonnenschein, J.~\"Ayst\"o, Phys.~Rev.~C \textbf{82,} (2010) 052501.
 \bibitem{Weber2008} C.~Weber, V.-V.~Elomaa, R.~Ferrer, C.~Fr\"ohlich, D.~Ackermann, J.~\"Ayst\"o, G.~Audi, L.~Batist, K.~Blaum, M.~Block, A.~Chaudhuri, M.~Dworschak, S.~Eliseev, T.~Eronen, U.~Hager, J.~Hakala, F.~Herfurth, F.~P.~He\ss{}berger, S.~Hofmann, A.~Jokinen, A.~Kankainen, H.~J.~Kluge, K.~Langanke, A.~Mart\'\i{}n, G.~Mart\'\i{}nez-Pinedo, M.~Mazzocco, I.~D.~Moore, J.~B.~Neumayr, Yu.~N.~Novikov, H.~Penttil\"a, W.~R.~Pla\ss{}, A.~V.~Popov, S.~Rahaman, T.~Rauscher, C.~Rauth, J.~Rissanen, D.~Rodr\'\i{}guez, A.~Saastamoinen, C.~Scheidenberger, L.~Schweikhard, D.~M.~Seliverstov, T.~Sonoda, F.-K.~Thielemann, P.~G.~Thirolf, G.~K.~Vorobjev, Phys.~Rev.~C \textbf{78,} (2008) 054310.
 \bibitem{Baruah2008}S.~Baruah, G.~Audi, K.~Blaum, M.~Dworschak, S.~George, C.~Gu\'enaut, U.~Hager, F.~Herfurth, A.~Herlert, A.~Kellerbauer, H.-J.~Kluge, D.~Lunney, H.~Schatz, L.~Schweikhard, C.~Yazidjian, Phys.~Rev.~Lett.~\textbf{101,} (2008) 262501.
 \bibitem{Dworschak2008}M.~Dworschak, G.~Audi, K.~Blaum, P.~Delahaye, S.~George, U.~Hager, F.~Herfurth, A.~Herlert, A.~Kellerbauer, H.-J.~Kluge, D.~Lunney, L.~Schweikhard, C.~Yazidjian, Phys.~Rev.~Lett.~\textbf{100} (2008) 072501.
 \bibitem{Elomaa2009}V.-V.~Elomaa, G.~K.~Vorobjev, A.~Kankainen, L.~Batist, S.~Eliseev, T.~Eronen, J.~Hakala, A.~Jokinen, I.~D.~Moore, Yu.~N.~Novikov, H.~Penttil\"a, A.~Popov, S.~Rahaman, J.~Rissanen, A.~Saastamoinen, H.~Schatz, D.~M.~Seliverstov, C.~Weber, J.~\"Ayst\"o, Phys.~Rev.~Lett.~\textbf{102,} (2009) 252501.
 \bibitem{Hardy2009}J.~C.~Hardy, I.~S.~Tower, Phys.~Rev.~C \textbf{79,} (2009) 055502.
 \bibitem{Kellerbauer2004}A.~Kellerbauer, G.~Audi, D.~Beck, K.~Blaum, G.~Bollen, B.~A.~Brown, P.~Delahaye, C.~Gu\'enaut, F.~Herfurth, H.~J.~Kluge, D.~Lunney, S.~Schwarz, L.~Schweikhard, C.~Yazidjian, Phys.~Rev.~Lett.~\textbf{93,} (2004) 072502.
 \bibitem{Mukherjee2004}M.~Mukherjee, A.~Kellerbauer, D.~Beck, K.~Blaum, G.~Bollen, F.~Carrel, P.~Delahaye, J.~Dilling, S.~George, C.~Gu\'enaut, F.~Herfurth, A.~Herlert, H.-J.~Kluge, U.~K\"oster, D.~Lunney, S.~Schwarz, L.~Schweikhard, C.~Yazidjian, Phys.~Rev.~Lett.~\textbf{93,} (2004) 150801.
 \bibitem{Savard2005}G.~Savard, F.~Buchinger, J.~A.~Clark, J.~E.~Crawford, S.~Gulick, J.~C.~Hardy, A.~A.~Hecht, J.~K.~P.~Lee, A.~F.~Levand, N.~D.~Scielzo, H.~Sharma, K.~S.~Sharma, I.~Tanihata, A.~C.~C.~Villari, Y.~Wang, Phys.~Rev.~Lett.~\textbf{95,} (2005) 102501.
 \bibitem{Bollen2006}G.~Bollen, D.~Davies, M.~Facina, J.~Huikari, E.~Kwan, P.~A.~Lofy, D.~J.~Morrissey, A.~Prinke, R.~Ringle, J.~Savory, P.~Schury, S.~Schwarz, C.~Sumithrarachchi, T.~Sun, L.~Weissman, Phys.~Rev.~Lett.~\textbf{96,} (2006) 152501.
 \bibitem{Eronen2006}T.~Eronen, V.~Elomaa, U.~Hager, J.~Hakala, A.~Jokinen, A.~Kankainen, I.~Moore, H.~Penttil\"a, S.~Rahaman, J.~Rissanen, A.~Saastamoinen, T.~Sonoda, J.~\"Ayst\"o, J.~C.~Hardy, V.~S.~Kolhinen, Phys.~Rev.~Lett.~\textbf{97,} (2006) 232501.
 \bibitem{George2007}S.~George, S.~Baruah, B.~Blank, K.~Blaum, M.~Breitenfeldt, U.~Hager, F.~Herfurth, A.~Herlert, A.~Kellerbauer, H.-J.~Kluge, M.~Kretzschmar, D.~Lunney, R.~Savreux, S.~Schwarz, L.~Schweikhard, C.~Yazidjian, Phys.~Rev.~Lett.~\textbf{98,} (2007) 162501.
 \bibitem{Eronen2008}T.~Eronen, V.-V.~Elomaa, U.~Hager, J.~Hakala, J.~C.~Hardy, A.~Jokinen, A.~Kankainen, I.~D.~Moore, H.~Penttil\"a, S.~Rahaman, S.~Rinta-Antila, J.~Rissanen, A.~Saastamoinen, T.~Sonoda, C.~Weber, J.~\"Ayst\"o, Phys.~Rev.~Lett.~\textbf{100,} (2008) 132502.
 \bibitem{Eronen2009}T.~Eronen, V.~V.~Elomaa, J.~Hakala, J.~C.~Hardy, A.~Jokinen, I.~D.~Moore, M.~Reponen, J.~Rissanen, A.~Saastamoinen, C.~Weber, J.~\"Ayst\"o, Phys.~Rev.~Lett.~\textbf{103,} (2009) 252501.
 \bibitem{Gabrielse1999}G.~Gabrielse, A.~Khabbaz, D.~S.~Hall, C.~Heimann, H.~Kalinowsky, W.~Jhe, Phys.~Rev.~Lett.~\textbf{82,} (1999) 3198.
 \bibitem{Shabaev2006}V.~M.~Shabaev, O.~V.~Andreev, A.~N.~Artemyev, S.~S.~Baturin, A.~A.~Elizarov, Y.~S.~Kozhedub, N.~S.~Oreshkina, I.~I.~Tupitsyn, V.~A.~Yerokhin, O.~M.~Zherebtsov, Int.~J.~Mass Spectrom.~\textbf{251,} (2006) 109.
 \bibitem{Stoehlker2008}Th.~St\"ohlker, A.~Gumberidze, M.~Trassinelli, V.~Andrianov, H.~F.~Beyer,
S.~Kraft-Bermuth, A.~Bleile, P.~Egelhof, and The FOCAL collaboration, Lect.~Notes Phys.~\textbf{745,} (2008) 157.
 \bibitem{Rainville2005}S.~Rainville, J.~K.~Thompson, E.~G.~Myers, J.~M.~Brown, M.~S.~Dewey, E.~G.~Kessler, R.~D.~Deslattes, H.~G.~B\"orner, M.~Jentschel, P.~Mutti, D.~E.~Pritchard, Nature \textbf{438,} (2005) 1096.
 \bibitem{VanDyck2004}R.~S.~Van Dyck, Jr., S.~L.~Zafonte, S.~Van Liew, D.~B.~Pinegar, P.~B.~Schwinberg, Phys.~Rev.~Lett.~\textbf{92,} (2004) 220802.
 \bibitem{VanDyck2006}R.~S.~Van Dyck, Jr., D.~B.~Pinegar, S.~Van Liew, S.~L.~Zafonte, Int.~J.~Mass Spectrom.~\textbf{251,} (2006) 231.
 \bibitem{Rainville2004}S.~Rainville, J.~K.~Thompson, D.~E.~Pritchard, Science \textbf{303,} (2004) 334.
 \bibitem{Shi2005}W.~Shi, M.~Redshaw, E.~G.~Myers, Phys.~Rev.~A \textbf{72,} (2005) 022510.
 \bibitem{Redshaw2008}M.~Redshaw, J.~McDaniel, E.~G.~Myers, Phys.~Rev.~Lett.~\textbf{100,} (2008) 093002.
 \bibitem{Bergstroem2002}I.~Bergstr\"om, C.~Carlberg, T.~Fritioff, G.~Douysset, J.~Sch\"onfelder, R.~Schuch, Nucl.~Instr.~Meth.~Phys.~Res.~A \textbf{487,} (2002) 618.
 \bibitem{Bergstroem2003}I.~Bergstr\"om, M.~Bj\"orkhage, K.~Blaum, H.~Bluhme, T.~Fritioff, Sz.~Nagy, R.~Schuch, Eur.~Phys.~J.~D \textbf{22,} (2003) 41.
 \bibitem{Dilling2006}J.~Dilling, R.~Baartman, P.~Bricault, M.~Brodeur, L.~Blomeley, F.~Buchinger, J.~Crawford, J.~R.~Crespo L\'opez-Urrutia, P.~Delheij, M.~Froese, G.~P.~Gwinner, Z.~Ke, J.~K.~P.~Lee, R.~B.~Moore, V.~Ryjkov, G.~Sikler, M.~Smith, J.~Ullrich, J. Vaz, Int.~J.~Mass Spectrom.~\textbf{251,} (2006) 198.
 \bibitem{Graeff1980} G.~Gr\"aff, H.~Kalinowsky, J.~Traut, Z.~Phys.~A \textbf{297,} (1980) 35.
  \bibitem{Wineland1975}D.~J.~Wineland, H.~G.~Dehmelt, J.~Appl.~Phys.~\textbf{46,} (1975) 919.
 \bibitem{Eliseev2011}S.~Eliseev, C.~Roux, K.~Blaum, M.~Block, C.~Droese, F.~Herfurth, H.-J.~Kluge, M.~I.~Krivoruchenko,~Yu.N. Novikov, E.~Minaya Ramirez, L.~Schweikhard, V.~M.~Shabaev, F.~$\check{\mathrm{S}}$imkovic, I.~I.~Tupitsyn, K.~Zuber, N.~A.~Zubova, Phys.~Rev.~Lett.~\textbf{106,} (2011) 052504.
 \bibitem{Rujula1982}R\'ujula, M.~Lusignoli, Phys.~Rev.~Lett.~B \textbf{118,} (1982) 429.
 \bibitem{Ferri2009}E.~Ferri, C.~Arnaboldi, G.~Ceruti, C.~Kilbourne, S.~Kraft-Bermuth, A.~Nucciotti, G.~Pessina, D.~Schaeffer, AIP Conference Proceedings \textbf{1185,} (2009).
 \bibitem{Otten2008}E.~W.~Otten, C.~Weinheimer, Rep.~Prog.~Phys.~\textbf{71,} (2008) 086201.
 \bibitem{Stoehlker2006}Th.~St\"ohlker, H.~F.~Beyer, A.~Gumberidze, A.~Kumar, D.~Liesen, R.~Reuschl, U.~Spillmann, M.~Trassinelli, Hyperfine Int.~\textbf{172,} (2006) 135.
 \bibitem{Brown1986}L.~S.~Brown, G.~Gabrielse, Rev.~Mod.~Phys.~\textbf{58,} (1986) 233.
 \bibitem{Gabrielse1989}G.~Gabrielse, L.~Haarsma, S.~L.~Rolston, Int.~J.~Mass Spectrom.~Ion Process.~\textbf{88,} (1989) 319.
 \bibitem{Gabrielse2009}G.~Gabrielse, Int.~J.~Mass Spectrom.~\textbf{279,} (2009) 107.
 \bibitem{Liu2010}Y.~Liua, M.~Hobein, A.~Solders, M.~Suhonen, R.~Schuch, Int.~J.~Mass Spectrom.~\textbf{294,} (2010) 28.
 \bibitem{Droese2011}C.~Droese, M.~Block, M.~Dworschak, S.~Eliseev, E.~Minaya Ramirez, D.~Nesterenko, L.~Schweikhard, Nucl.~Instrum.~Methods A \textbf{632,} (2011) 157.
  \bibitem{VanDyck1999}R.~S.~Van Dyck, Jr., D.~L.~Farnham, S.~L.~Zafonte, P.~B.~Schwinberg, Rev.~Sci.~Instrum.~\textbf{70,} (1999) 1665.
 \bibitem{Kluge2008}H.-J.~Kluge, T.~Beier, K.~Blaum, L.~Dahl, S.~Eliseev, F.~Herfurth, B.~Hofmann, O.~Kester, S.~Koszudowski, C.~Kozhuharov, G.~Maero, W.~N\"ortersh\"auser, J.~Pfister, W.~Quint, U.~Ratzinger, A.~Schempp, R.~Schuch, Th.~St\"ohlker, R.C.~Thompson, M.~Vogel, G.~Vorobjev, D.F.A.~Winters and G.~Werth, Adv.~Quantum Chem.~\textbf{53,} (2008) 83.
 \bibitem{Levine1988}M.~Levine, R.~Marrs, J.~Henderson, D.~Knapp, M.~Schneider, Physica Scripta, \textbf{T22,} (1988) 157.
 \bibitem{Levine1989}M.~Levine, R.E.~Marrs, J.N.~Bardsley, P.~Beiersdorfer, C.L.~Bennett, M.H.~Chen, T.~Cowan, D.~Dietrich, J.R.~Henderson, D.A.~Knapp, A.~Osterheld, B.M.~Penetrante, M.B.~Schneider, J.H.~Scofield, Nucl.~Instrum.~Methods B \textbf{43,} (1989) 431.
 \bibitem{Marrs1994a}R.~E.~Marrs, S.~R.~Elliott, D.~A.~Knapp, Phys.~Rev.~Lett.~\textbf{72,} (1994) 4082.
 \bibitem{Marrs1994b}R.~E.~Marrs, P.~Beiersdorfer, D.~Schneider, Physics Today \textbf{47,} (1994) 27.
 \bibitem{Beiersdorfer1997}P.~Beiersdorfer, S.~R.~Elliott, J.~R.~Crespo L\'opez-Urrutia, K.~Widmann, Nucl.~Phys.~A \textbf{626,} (1997) 357.
 \bibitem{Ovsyannikov1999}V.~P.~Ovsyannikov, G.~Zschornack, Rev.~Sci.~Instrum.~\textbf{70,} (1999) 2646.
 \bibitem{Zschornack2010}G.~Zschornack, M.~Kreller, A.~Silze, VP.~Ovsyannikov, F.~Grossmann, R.~Heller, U.~Kentsch, M.~Schmidt, A.~Schwan, F.~Ullmann, Rev.~Sci.~Instrum.~\textbf{81,} (2010) 02A507.
 \bibitem{Dreebit2011}See: http://www.dreebit.com.
 \bibitem{Crespo2000}J.~R.~Crespo L\'opez-Urrutia, B.~Bapat, B.~Feuerstein, A.~Werdich, J.~Ullrich, Hyperfine Int.~\textbf{127,} (2000) 497.
 \bibitem{Crespo2001}J.~R.~Crespo L\'opez-Urrutia, B.~Bapat, I.~Draganic, A.~Werdich, J.~Ullrich, Physica Scripta \textbf{T92,} (2001) 110.
 \bibitem{Crespo2004}J.~R.~Crespo L\'opez-Urrutia, J.~Braun, G.~Brenner, H.~Bruhns, A.~Lapierre, A.~J.~Gonz\'alez Mart\'inez, V.~Mironov, R.~Soria Orts, H.~Tawara, M.~Trinczek, and J.~Ullrich, Rev.~Sci.~Instrum.~\textbf{75,} (2004) 1560.
 \bibitem{Martinez2005}A.~J.~Gonz\'alez Mart\'inez, J.~R.~Crespo L\'opez-Urrutia, J.~Braun, G.~Brenner, H.~Bruhns, A.~Lapierre, V.~Mironov, R.~Soria Orts, H.~Tawara, M.~Trinczek, J.~Ullrich, J.~H.~Scofield, Phys.~Rev.~Lett.~\textbf{94,} (2005) 203201.
 \bibitem{Manura2007}D.~J.~Manura, D.~A.~Dahl, \textit{SIMION$^{TM}$ Version 8.0 User Manual} (Scientific Instrument Services, Inc., Ringoes, 2007)
 \bibitem{Stahl2005}S.~Stahl, J.~Alonso, S.~Djekic, H.-J.~Kluge, W.~Quint, J.~Verdu, M.~Vogel, G.~Werth, J. Phys. B \textbf{38,} (2005) 297.
 \bibitem{Sturm2011a}S.~Sturm, A.~Wagner, B.~Schabinger, K.~Blaum, accepted by Phys.~Rev.~Lett.~(2011).
 \bibitem{Allan1966}D.~Allan, Proceedings of the IEEE \textbf{54,} (1996) 221.
 \bibitem{AlaVAR52} See: http://www.alamath.com/ for a program to calculate the Allan deviation.
 \bibitem{Verdu2008} J.~Verd\'u, S.~Kreim, K.~Blaum, H.~Kracke, W.~Quint, S.~Ulmer, J.~Walz, New J.~Phys.~\textbf{10,} (2008) 103009
 \bibitem{Gabrielse1984} G.~Gabrielse, F.~C.~MacKintosh, Int.~J.~Mass Spectrom.~\textbf{57,} (1984) 1.
 \bibitem{Roux2011} C.~Roux, Ch.~B\"ohm, A.~D\"orr, S.~Eliseev, M.~Goncharov, Yu.~Novikov, J.~Repp, S.~Sturm, S.~Ulmer, K.~Blaum, Appl.~Phys.~B, to be published (2011)
 \bibitem{Jefferts1993}S.~R.~Jefferts, T.~Heavner, P.~Hayes, G.~H.~Dunn, Rev.~Sci.~Instrum.~\textbf{64,} (1993) 737.
 \bibitem{Ulmer2009}S.~Ulmer, H.~Kracke, K.~Blaum, S.~Kreim, A.~Mooser, W.~Quint, C.~C.~Rodegheri, J.~Walz, Rev.~Sci.~Instrum.~\textbf{80,} (2009) 123302.
 \bibitem{Cornell1990}E.~A.~Cornell, R.~M.~Weisskoff, K.~R.~Boyce, D.~E.~Pritchard, Phys.~Rev.~A \textbf{41,} (1990) 312.
 \bibitem{Cornell1989}E.~A.~Cornell, R.~M.~Weisskoff, K.~R.~Boyce, R.~W.~Flanagan, G.~P.~Lafyatis, D.~E.~Pritchard, Phys.~Rev.~Lett.~\textbf{63,} (1989) 1674.
 \bibitem{Beck2003}D.~Beck, H.~Brand, GSI Scientific Rep.~\textbf{2002,} (2003) 210.
 \bibitem{Beck2004}D.~Beck, K.~Blaum, H.~Brand, F.~Herfurth, S.~Schwarz, Nucl.~Instrum.~Methods A \textbf{527,} (2004) 567.
\end{thebibliography}
\end{document}